\documentclass[aps,prd,twocolumn,showpacs,superscriptaddress,nofootinbib,10pt]{revtex4-2}

\usepackage{acronym}
\usepackage{amsfonts}
\usepackage{amsmath}
\usepackage{amssymb}
\usepackage{bm}
\usepackage{cases}
\usepackage{color}
\usepackage{comment}
\usepackage[dvipsnames]{xcolor}
\usepackage{enumerate}
\usepackage{epsfig}
\usepackage{etoolbox}
\usepackage{fancyref}
\usepackage{fancyhdr}
\usepackage[T1]{fontenc} 
\usepackage{graphicx}
\usepackage{hyperref}
\usepackage{indentfirst}
\usepackage{latexsym}
\usepackage{mathrsfs}
\usepackage{mciteplus}
\usepackage{multirow}
\usepackage{rotating}
\usepackage{scrextend}
\usepackage{soul}
\usepackage{subfigure}
\usepackage{verbatim}
\usepackage{amsthm} 

\newcommand{\be}{\begin{equation}}
\newcommand{\ee}{\end{equation}}
\newcommand{\bea}{\begin{eqnarray}}
\newcommand{\eea}{\end{eqnarray}}

\font\tenscr=rsfs10 scaled1100
\font\sevenscr=rsfs7 
\font\fivescr=rsfs5 
\skewchar\tenscr='177
\skewchar\sevenscr='177
\skewchar\fivescr='177
\newfam\scrfam
\textfont\scrfam=\tenscr
\scriptfont\scrfam=\sevenscr
\scriptscriptfont\scrfam=\fivescr

\def\scri{{\fam\scrfam I}}

\begin{document}

\title[Integrability in Perturbed Black Holes: Background Hidden Structures]{Integrability in Perturbed Black Holes: Background Hidden Structures}

\author{Jos\'e Luis Jaramillo}
\email{Jose-Luis.Jaramillo@u-bourgogne.fr}
\affiliation{Institut de Math\'ematiques de Bourgogne  (IMB), UMR 5584,  CNRS, Universit\'e  de  Bourgogne,  F-21000  Dijon,  France}

\author{Michele Lenzi}
\email{lenzi@ice.csic.es}
\affiliation{Institut de Ci\`encies de l'Espai (ICE, CSIC), Campus UAB, Carrer de Can Magrans s/n, 08193 Cerdanyola del Vall\`es, Spain}
\affiliation{Institut d'Estudis Espacials de Catalunya (IEEC), Carrer Esteve Terradas, 1, Edifici RDIT, Campus PMT-UPC, 08860 Castelldefels, Spain}

\author{Carlos F. Sopuerta}
\email{carlos.f.sopuerta@csic.es}
\affiliation{Institut de Ci\`encies de l'Espai (ICE, CSIC), Campus UAB, Carrer de Can Magrans s/n, 08193 Cerdanyola del Vall\`es, Spain}
\affiliation{Institut d'Estudis Espacials de Catalunya (IEEC), Carrer Esteve Terradas, 1, Edifici RDIT, Campus PMT-UPC, 08860 Castelldefels, Spain}

\date{\today}

\begin{abstract}
 
In this work we investigate the presence of integrable hidden structures in the dynamics of perturbed non-rotating black holes (BHs). 
This can also be considered as a first step in a wider program of an effective identification of “slow” and “fast” degrees of freedom (DoFs) in the (binary) BH dynamics, following a wave-mean flow perspective. The slow DoFs would be associated with a nonlinear integrable dynamics, on which the fast ones propagate following an effective linear dynamics.
BH perturbation theory offers a natural ground to test these properties. Indeed, the decoupling of Einstein equations into wave master equations with a potential provides an instance of such splitting into (frozen) slow DoFs (background potential) over which the linear dynamics of the fast ones (perturbation master functions) evolve.
It has been recently shown that these wave equations possess an infinite number of symmetries that correspond to the flow of the infinite hierarchy of Korteweg-de Vries (KdV) equations. Starting from these results, we systematically investigate the presence of integrable structures in BH perturbation theory. 
We first study them in Cauchy slices and then extend the analysis to hyperboloidal foliations. This second step introduces a splitting of the master equation into bulk and boundary contributions, unveiling an underlying structural relation with the slow and fast DoFs. This insight represents a first step to establish the integrable structures associated to the slow DoFs as bulk symmetries of the dynamics of perturbed BHs.

\end{abstract}

\maketitle

\section{Introduction: Hidden symmetries in black hole dynamics}
\label{s:intro}

The general setting of the present work is to progress in the exploration of the structurally stable qualitative features of black hole (BH) dynamics in the strong field regime, by identifying a set of background (hidden symmetry) structures underlying such stable dynamical features. 

{Our focus is placed on a class of structurally stable aspects of gravitational dynamics, that is, on features valid for generic (non-fine tuned) initial data configurations. Such kind of dynamics is not the only relevant one in general relativity (GR), as the discovery of critical phenomena in gravitational collapse~\cite{choptuik1993universality} has plainly demonstrated and is further endorsed from a ``dynamical systems'' approach to GR (see e.g.~\cite{gundlach2007critical,fischer1979initial}). Under such a dynamical systems system perspective our focus would be restrained to a particular ``basin of attraction''.

A specially important instance of a structurally stable dynamical feature in our context is given by the so-called \emph{soliton resolution conjecture}~\cite{bizon2022characteristic}, according to which generic 
global-in-time (general relativistic) non-linear wave dynamics would decouple universally at late times into ``soliton’' solutions plus ``radiation''.
A dramatic example of such soliton resolution behaviour in GR  would be provided by the so-called ``establishment picture of gravitational collapse'', namely a sequence of theorems and conjectures that provide a universal and structural stable picture in which generic  gravitationally collapsing configurations asymptotically approach a universal class of BH spacetimes (namely sub-critical Kerr, the soliton-like solution in this setting) plus the (gravitational) radiation emitted at null infinity and through the horizon.

More broadly, we aim at exploring the role of such a class of structurally stable features of the theory in the setting of general BH  dynamics, including binary BH (BBH) mergers. The \emph{soliton resolution conjecture} properly applies to non-linear {\em dispersive} equations \cite{tao2009solitons}. In this setting, as a methodological  approach, we adopt the perspective recently proposed in \cite{Jaramillo:2023day}  (see also \cite{Jaramillo:2022oqn,Jaramillo:2022kuv,Jaramillo:2024nvr}). Such a methodology relies on two elements, on the one hand a ``wave-mean flow'' approach to the dynamics ~\footnote{More specifically, this corresponds to the layer discussed in Sec.~V of the hierarchical proposal to BBH dynamics in Ref.~\cite{Jaramillo:2023day} (see Table 1 in that reference and also Refs.~\cite{Jaramillo:2022mkh,Jaramillo:2022oqn,Jaramillo:2024nvr}).} and, on the other hand, a so-called ``asymptotic reasoning'' strategy~\cite{batterman2002devil}. The former proposes the effective separation of the dynamics into ``fast'' and ``slow'' degrees of freedom (DoFs), whereas the latter filters away some DoFs of the theory thus sacrificing precision and exactness in order to unveil the underlying (universal) structurally stable patterns, typically in an asymptotic limit of a relevant physical parameter.

\subsection{Wave-mean flow and asymptotic reasoning}
\label{s:wave_mean-flow_asymp-reason}

To be more concrete, and also with the aim of identifying the mechanisms underlying the structural stable features of BH strong field dynamics, 
in particular the already mentioned \emph{soliton resolution conjecture} in the BBH context, we adopt an effective description of the gravitational dynamics with separation into background ``slow'' and scattered ``fast'' DoFs. The slow DoFs are proposed to evolve according to dynamics given in terms of non-linear dispersive partial differential equations (PDEs), whereas the fast DoFs would propagate subject to effective linear dynamics on the background provided  by the (non-linear) slow DoFs. That is:
\begin{itemize}
    \item[a)] Slow DoFs: Subject to effective dispersive nonlinear PDEs, possibly integrable or (quasi-)integrable, and providing the background for the fast DoFs.

    \item[b)]  Fast DoFs: Subject to effective linear wave dynamics, propagating on the slowly dynamical background. In non-linear theories as GR they induce a backreaction on the background. 
\end{itemize}
In a second stage, we consider a further hierarchy of dynamical levels in such a fast-slow ``wave-mean flow'' approach, by zooming into specific aspects of the dynamics and freezing some of the DoFs. This is in the spirit of the aforementioned  asymptotic reasoning where some of the DoFs are filtered out in an attempt of unveiling the structurally stable aspects of a given phenomenon. Specifically, we consider the following hierarchy:
\begin{itemize}
\item[i)] Full ``slow-fast'' dynamics \cite{Jaramillo:2023day,Jaramillo:2022oqn}: namely the ``wave-mean flow'' introduced above, possibly with back-reaction of the (linear) fast DoFs on the slow ones.

\item[ii)] Linear scattering of ``test fields'' on a
dynamical background: ``fast'' DoFs do not backreact on the dynamically slow background.

\item[iii)] Linear scattering on a stationary background: scattering off the background defined by the (frozen) slow DoFs. 
\end{itemize}
The systematic study of such a hierarchy defines a work program, which by itself constitutes a subprogram of the one described in Refs.~\cite{{Jaramillo:2023day,Jaramillo:2022oqn}} (see also \cite{Jaramillo:2024nvr}), each layer focusing on a particular dynamical aspect of the problem.
The main goal of this work is to focus on the identification of the relevant algebraic structures controlling the structurally stable aspects of the dynamics, more concretely, on the specific ones of the bulk background spacetime (slow DoFs). Pushing a bit further the asymptotic reasoning scheme, the Ansatz is that such algebraic structures, meant to be present and controlling the whole hierarchy, are more easily identifiable in the lowest layer ---point iii) above. In this article, we focus on the linear scattering on a stationary background (the BH background) as a first step in a further reaching program.

\subsection{Background structures in linear scattering}
\label{s:background_structures}
The study of the hidden symmetries in the spirit that we discuss here, and in the setting of BH Perturbation Theory (BHPT), goes back to the work of Chandrasekhar and collaborators (see~~\cite{Chandrasekhar:1992bo} for a comprehensive review). In particular, crucial insight into the BH QuasiNormal Mode (QNM) isospectrality of the Schwarzschild BH is gained by the unveiling of a set of conserved quantities associated with an underlying Korteweg-De Vries (KdV) structure~\cite{Chandrasekhar:1992bo} and the existence of Darboux transformations between the potentials and master functions of the two different parity modes (i.e. axial/odd-parity and polar/even-parity modes)~\cite{Glampedakis:2017rar}. A systematic study of such KdV and Darboux `hidden symmetries' has been developed in a recent series of articles~\cite{Lenzi:2021wpc,Lenzi:2021njy,Lenzi:2022wjv,Lenzi:2023inn,Lenzi:2024tgk}. In the following, we elaborate further on such recent results and on the structure of BHPT, giving more details and results.

In Ref.~\cite{Lenzi:2021wpc}, the full space of master functions and equations for the perturbations of Schwarzschild BHs was constructed under very general assumptions. Two branches of master equations and functions were distinguished: (i) The {\em standard} branch, which essentially contains master equations with the known BH potentials, namely the Regge-Wheeler potential~\cite{Regge:1957td} for odd-parity perturbations and the Zerilli potential~\cite{Zerilli:1970la} for even-parity ones; and (ii) the {\em Darboux} branch, where there is an infinite set of master equations with new BH potentials.  In Ref.~\cite{Lenzi:2021njy}, it was found that all these master equations and functions can be linked by means of Darboux transformations (see also~\cite{Glampedakis:2017rar}), establishing in this way their physical equivalence in the sense that the Darboux transformation preserves the spectrum, that is, the different problems associated with the different time-independent master equations are isospectral. More generally, all these master equations lead to the same spectrum of QNMs and also to the same reflection and transmission coefficients for scattering states, thus extending the result of Refs.~\cite{1980RSPSA.369..425C,Chandrasekhar:1992bo}. Therefore, Darboux transformations emerge as a symmetry of the space of master equations and functions describing the perturbations of the Schwarzschild BH geometry, and for 
 this reason, this property has been named as {\em Darboux covariance}~\cite{Lenzi:2021njy} (see Ref.~\cite{Lenzi:2024tgk} for the extension to the presence of perturbative sources).

Moreover, the structure of the space of master functions and equations gets enriched by the introduction~\cite{Lenzi:2021njy} of inverse scattering techniques~\cite{Faddeev:1959yc, Faddeev:1976xar, Deift:1979dt, Novikov:1984id, Ablowitz:1981jq}. These techniques can be used to solve certain non-linear evolution problems~\cite{Miura:1968JMP.....9.1202M,Gardner:1967wc} by establishing a mathematical connection with \emph{linear} wave scattering problems. Those non-linear PDEs that can be solved by these methods are then characterised as {\em integrable}. The Korteweg-de Vries (KdV) equation, which describes a very wide range of physical phenomena~\cite{doi:10.1137/1018076}, provides a paradigmatic example. The key idea behind the application of the inverse scattering method to this integrable PDE setting is to cast the unknown satisfying the non-linear evolution PDE as a time-evolving potential that can be associated to a (linear) scattering problem. That is, the unknown can then be reconstructed from the (non-linear) evolution of the associated scattering data. The scattering data is obtained from the associated linear, and time-independent, problem. In the case of the KdV equation, the linear problem is the time-independent Schr\"odinger equation. Another remarkable result is the appearance of an infinite series of conservation laws for the KdV equation with the corresponding set of conserved quantities~\cite{Miura:1968JMP.....9.1204M, Zakharov:1971faa, Lax:1968fm}, the so-called \emph{KdV integrals}. We can apply these techniques to our master equations in the frequency domain and generate the associated set of KdV integrals~\cite{Lenzi:2021njy}. One can see that the KdV equation constitutes an isospectral deformation of the master equations, in such a way that the transmission coefficients and the QNMs are preserved by the KdV deformation.  Crucially, it was shown in~\cite{Lenzi:2021njy} that the KdV integrals are invariant under Darboux transformations. That is, they are the same for all the potentials in the set of master equations.

~

In Refs.~\cite{Lenzi:2022wjv,Lenzi:2023inn}, it has been shown that the KdV integrals associated with the BH effective potential fully determine the transmission probability, or greybody factors, which contains all the relevant physical information about BH scattering processes. The way in which the KdV integrals determine the transmission probability is through a {\em moment problem} (see, e.g.~\cite{Shohat1943ThePO,akhiezer1965classical,schmudgen2017moment}). That is, the KdV integrals turn out to be the moments (up to a trivial multiplicative factor) of a distribution function associated with the transmission probability. This moment problem is determinate, so that a solution exists and it is unique. This means that it is possible to invert the moment problem to find the greybody factors~\cite{Lenzi:2022wjv,Lenzi:2023inn}. 

\subsection{Hyperboloidal approach to BHPT}
\label{s:hyperboloidal_approach}
The previous discussion focuses on underlying ``bulk'' algebraic structures in BHPT. However, another class of relevant algebraic structures in this BH scattering problem is associated with spacetime asymptotics, namely the far wave zone for massless fields, modelled by null infinity $\scri^+$. Such scattering algebraic structures at infinity are more difficult to grasp in the scheme presented above, the reason laying in the use of Cauchy slices that reach spatial infinity $i^0$ rather than $\scri^+$. An approach in principle facilitating addressing both bulk and spacetime asymptotic structures is given by the use of the so-called hyperboloidal foliations. Indeed such a scheme permits, on the one hand, to fully upgrade the bulk analysis developed in the Cauchy formulation and, on the other hand, to cast the scattering problem in a framework permitting to make explicit the relevant structures at null infinity.

More specifically, the hyperboloidal scheme casts the evolution problem in terms of a hyperboloidal foliation $\{\Sigma_\tau\}$ of spacetime, where slices $\Sigma_\tau$ are spacelike hypersurfaces that smoothly transversally intersect null infinity $\scri^+$ and, in the case of BH spacetimes, also the event horizon. In this manner such slices, and in particular the ``hyperboloidal time'' function $\tau$, permit to explicitly parametrize the relevant geometry at $\scri^+$ and the BH horizon. This scheme provides a kind of geometric interpolation between Cauchy and null slicing formulations of dynamics, naturally leading to a framework adapted both to the discussion of bulk structures in a methodology akin to that 
employed in the Cauchy approach and, simultaneously, to the discussion of spacetime asymptotic ``boundary'' structures. Connecting with this latter point, the question about the outgoing boundary conditions defining our scattering problem is of particular importance in our discussion. The hyperboloidal scheme provides an in-built geometric enforcement of such outgoing boundary conditions, due to the null (in the asymptotically flat case) nature of the considered spacetime boundaries, $\scri^+$ and the event horizon: characteristics follow the null cones and are automatically outward-pointing at the boundaries, imposing outgoing boundary conditions on physical fields such as the master functions.

The use of hyperboloidal foliations has proved very successful in its application to both fundamental and applied problems in spacetime dynamics and BH scattering. It provides one of the main ingredients in key results such as the semi-global stability of Minkowski by Friedrich \cite{friedrich1986existence}, BH stability results in cosmological scenarios~\cite{vasy2013microlocal,hafner2021linear,Hintz:2016gwb}, the characterisation of QNMs as eigenvalues of a non-selfadjoint operator~\cite{Warnick:2013hba,Ansorg:2016ztf}, the numerical stability of the exploration of solution resolution in \cite{bizon2022characteristic}, or the analysis of the spectral instability of BH QNMs~\cite{Jaramillo:2020tuu,Jaramillo:2021tmt,Warnick:2024usx,Boyanov:2024fgc}. In recent years hyperbolodial foliations have become one of the main tools for the theoretical and numerical analysis of compact object sources of gravitational waves (cf. excellent discussions in \cite{Zenginoglu:2007jw,Zenginoglu:2011jz,panosso2024hyperboloidal,Zenginoglu:2024bzs,Hyperboloid_webpage,PanossoMacedo:2024nkw}).

A most important aspect of the hyperboloidal scheme in our present setting, namely focusing on the identification of the relevant algebraic structures in BH scattering, is that it provides a neat separation of bulk and asymptotic degrees of freedom. This feature manifests explicitly in the structure of the infinitesimal time generator of perturbations and it is the consequence of two ingredients: i) the choice of a 
``hyperboloidal time'' $\tau$ related to the ``Cauchy time'' $t$ through an appropriate height function (i.e. $\tau= t + h(x)$) introducing the hyperboloidal foliation $\{\Sigma_\tau\}$ to cast the dynamics, and ii) a compactification of the hyperboloidal hypersurfaces $\Sigma_\tau$. The first element adapts the treatment to the relevant outgoing asymptotics, whereas the second one recasts such outgoing boundary conditions in terms of the regularity of the scattered field at the boundaries (the latter is akin~\footnote{We thank J.A. Valiente-Kroon for insisting on this point. We refer to~\cite{juan_valiente-kroon} for a discussion on the adaptation of  Melrose's ``geometric scattering'' to the hyperboloidal approach to spacetime dynamics.} to the underlying philosophy in Melrose's `geometric scattering'~\cite{melrose1995geometric}). Enforcing points i) and ii) casts, upon a first-order reduction in time, the wave dynamics of the master function $\phi$ in Schrödinger form 
\bea
\label{e:wave_eq_1storder_intro}
 \partial^{}_\tau u = i\,\mathbb{L} u \ \ , \ \ u =
\begin{pmatrix}
  \phi \\ \psi=\partial^{}_\tau \phi
\end{pmatrix}  \ ,
\eea
with the operator $\mathbb{L}$ given by
\bea
\label{e:L_operator_intro}
\mathbb{L} =\frac{1}{i}\!  \left(
  \begin{array}{c|c}
    0 & 1 \\ \hline {\cal L}^{}_1 & {\cal L}^{}_2
  \end{array}
  \right).
  \eea
The structure of $\mathbb{L}$ reflects the above-commented separation into bulk and asymptotic structures, respectively encoded in the ${\cal L}_1$ and ${\cal L}_2$ parts. On the one hand, the operator ${\cal L}_1$ stands 
as the structural counterpart in the hyperboloidal scheme of the Schr\"odinger operator in the Cauchy slicing setting of section \ref{s:background_structures}. Specifically, ${\cal L}_1$ is a (singular) Sturm-Liouville operator, namely a second-order elliptic operator that includes the effective scattering potential and provides the elements for the discussion of the background structures in linear scattering (namely, Darboux covariance and KdV structures). Its ``singular'' character, namely the vanishing at the boundaries of its principal as a consequence of the (`Melrose's) compactification of $\Sigma_\tau$, results in the absence of explicit boundary conditions that are now encoded in the structure of the operator $\mathbb{L}$ itself. On the other hand, the operator ${\cal L}_2$ is the structural piece encoding the outgoing nature of the boundary conditions of the scattering problem. That is, if ${\cal L}_1$ incorporates the boundary conditions into $\mathbb{L}$, it is ${\cal L}_2$ the element that characterises precisely its outgoing nature. It can actually be shown~\cite{Jaramillo:2020tuu} that ${\cal L}_2$ is the term responsible of the non-self-adjoint character of $\mathbb{L}$ and that the corresponding dissipative character of the latter is precisely due to the flow of the scattered field through the spacetime boundaries, whose flux is controlled by the parameters in ${\cal L}_2$. Moreover, the term breaking the self-adjoint nature of the infinitesimal time generator, i.e. the operator $\mathbb{L}-\mathbb{L}^\dagger$ (where the adjoint $\mathbb{L}^\dagger$ is defined with respect to a natural `energy scalar product') has a distributional nature with support at null infinity and the BH horizon. In other words, the missing DoFs in the hyperboloidal non-self-adjoint formulation of the problem, associated with the loss of DoFs propagating from the bulk through the boundary, are sharply accounted by the flux controlled by the ${\cal L}_2$ term, the latter being indeed a structure essentially {\em living} on the boundaries. This has led to the proposal in~\cite{Gasperin:2021kfv} that one should be able to account for the loss of ``fast'' DoFs reaching the boundaries in terms of formal DoFs ``living'' on the spacetime boundaries, that would work as a kind of ``geiger counters'' of ``fast'' DoFs (their inclusion in the problem rendering it conservative, since the total account of energy is conserved by construction). More ambitiously, Ref.~\cite{Gasperin:2021kfv} proposes that asymptotic symmetries [namely, the Bondi-Metzner-Sachs (BMS) group of symmetries~\cite{Bondi:1962px,Sachs:1962wk}, or an appropriate extension] would stand as a dynamical symmetry generating the phase space of DoF on the boundary, namely the (asymptotic) ``fast'' DoFs.

\subsection{The ``KdV-Virasoro-Schwarzian derivative'' triangle: bulk structures in a BH dynamics program }
\label{s:KdV-Virasoro-Schwarzian}
The present article represents a first step into a more ambitious project aiming at identifying the algebraic structures giving support to the (effective) integrability of a sector of the BH dynamics in the strong regime under a ``wave-mean flow'' perspective. Under the (asymptotic reasoning) assumption that the analysis of linear perturbations (fast DoFs) on a stationary background (frozen slow DoFs) is enough to identify the relevant underlying structures, we have separated the problem into a bulk (slow) and asymptotic (fast) parts, by making use of a hyperboloidal scheme. Specifically, in the present work we focus on the study of the structures underlying the ${\cal L}_1$ bulk piece in the infinitesinal time generator $\mathbb{L}$, leaving the analysis of the ${\cal L}_2$ asymptotic piece for a further study. From a heuristic perspective, we expect the interplay between the bulk and the asymptotic parts to be realised in a semi-direct structure, with the bulk symmetries acting on the asymptotic (boundary) ones. This expectation justifies our methodological choice to split the problem into simpler non-trivial parts, focusing first on the bulk part (namely ${\cal L}_1$ in the hyperboloidal scheme).

Dwelling then in the bulk part, this work focuses on the bulk symmetries
and explores the heuristic intuition that the KdV structure of BHPT~\cite{Lenzi:2021njy,Lenzi:2022wjv} is related to conformal symmetries of the bulk DoFs. This leads to the identification of the role of the Virasoro algebra in this setting and that of the Schwarzian derivative in the transformation of the scattering potential under conformal mappings. In this setting, the ``KdV-Virasoro-Schwarzian derivative'' triangle is identified as providing the basic ingredients for the underlying bulk integrability (cf. \cite{Semenov_r-matrix_Approach} for a systematic discussion in a larger integrability setting).   
This is done first in the setting of a Cauchy description, where the KdV symmetries first appeared~\cite{Lenzi:2021njy,Lenzi:2022wjv}, then finding that such a ``bulk'' symmetry structure is (partially) preserved when passing to the hyperboloidal foliation. 

~

The plan of the article is the following. In Sec.~\ref{Ss:bhpt-summary} we first review the basic concepts and tools of BHPT and then introduce the complete landscape of the possible master equations decouplings allowed by Einstein equations. In particular, in subsection~\ref{Ss:darboux-covariance} we review the symmetry structure hiding behind such infinite hierarchy of master equations, namely Darboux covariance. In Sec.~\ref{Ss:bulk-cauchy} we explore the bulk symmetries, in Cauchy slices, associated to KdV integrable structures. First, we introduce the KdV isospectrality and some interesting relations to SUSY quantum mechanics. Then, we show the connection with the conformal transformations and the Virasoro algebra. We conclude this section showing how the conformal transformations act on the master equations. In Sec.~\ref{Ss:bulk-hyperboloidal} we present how the structures studied in the previous section can be carried to a hyperboloidal scheme. With this aim we review the hyperboloidal approach in BHPT and introduce a general transformation of the master function, showing the covariance of the hyperboloidal master equation under such transformation. Finally, we conclude with a specific choice of the transformation which provides the connection with the bulk symmetries investigated in Sec.~\ref{Ss:bulk-cauchy}.

~

Throughout this paper, otherwise stated, we use geometric units in which $G = c = 1\,$. For convenience, we use multiple notations for partial derivatives of a given function $f(x)$: $\partial f/\partial x$, $\partial_x f$, $f_{,x}$.

\section{Linear scattering on a stationary background: BHPT} \label{Ss:bhpt-summary}

In this section we set the stage for the program described in the Introduction (Sec.~\ref{s:intro}) by first reviewing the study of gravitational perturbations on a BH background~\cite{Chandrasekhar:1992bo} (see also~\cite{Nagar:2005ea,Martel:2005ir,Lenzi:2021wpc,Chaverra:2012bh}). We further summarize some recent developments~\cite{Lenzi:2021wpc, Lenzi:2021njy, Lenzi:2022wjv} which guide us into the exploration of the structurally stable (hidden) structures which we expect to appear once a combined approach of wave-mean flow and asymptotic reasoning is considered (see Sec.~\ref{s:wave_mean-flow_asymp-reason}). In doing so, we highlight the relevant features of BHPT which connect with the hierarchical program described by points i),ii),iii) in Sec.~\ref{s:wave_mean-flow_asymp-reason}. The perturbative scheme is convenient since it allows for a sharp distinction between background/“slow” DoFs and perturbative/“fast” ones, such that the full metric of the perturbed spacetime can be written as
\begin{equation}
g^{}_{\mu\nu} = \widehat{g}^{}_{\mu\nu} + h^{}_{\mu\nu} \,,
\label{fundamental-perturbative-equation}
\end{equation}
where  one assumes that $|h_{\mu\nu}|\ll|\widehat{g}_{\mu\nu}|$ (see, e.g.~\cite{Stewart:1974uz,Wald:1984cw} for a precise formulation of this statement). Here, the background metric $\widehat{g}_{\mu\nu}$ satisfies the full Einstein equations while the dynamics of the perturbations $h_{\mu\nu}$ is linear. Schematically, this can be written as follows
\begin{eqnarray}
\hat{G}[\hat{g}]^{}_{\mu\nu} &=& 0
\\
\delta G[h^{}_{}]^{}_{\mu\nu} &=& 0 \,, \label{first-order-eqs}
\end{eqnarray}
where $\delta G$ is the linearized Einstein operator. 
However, the identification of background/“slow” and perturbative/“fast” DoFs is not apparent at this stage, as we still have to identify the correct background quantities which are related to integrable/stable structures and the perturbative DoFs associated to the linear asymptotic dynamics. The perturbation theory framework guides us towards the decoupling of Einstein equations for the perturbations~\eqref{first-order-eqs}
into (infinite) master equations with a potential and a wave function which will be directly related to “slow” and “fast” DoFs respectively. This already reduces the problem to the study of linear scattering (of gravitational perturbations) on a stationary background~\footnote{We however only consider static spherically symmetric backgrounds in this work.}.

 \subsection{The frozen \emph{slow} DoFs: spherically symmetric background}

The physical system we study is that of a perturbed isolated BH which is radiating away some residual energy to reach a spherically symmetric equilibrium configuration. Therefore, the background spacetime is described by the Schwarzschild spacetime~\cite{Schwarzschild:1916uq}~\footnote{Most of these results can be straightforward extended to non-asymptotically flat spacetimes like Schwarzschild-de Sitter and Schwarzschild-Anti de Sitter (see Refs.\cite{Lenzi:2024tgk,Grozdanov:2023txs}).}
\begin{equation}
d\hat{s}{}^2=\widehat{g}^{}_{\mu\nu}dx^{\mu}dx^{\nu}=-f(r)\,dt^2+\frac{dr^2}{f(r)}+r^2d\Omega^2\,,
\label{schwarzschild-metric}
\end{equation}
where
\begin{equation}
f(r) = 1-\frac{r_{s}}{r}\,.  \label{fsch-expression}
\end{equation}
Here $r_s$ is the location of the event horizon, the Schwarzschild radius: $r_{s} = 2GM/c^{2}=2M\,$, with $M$ being the BH mass. Due to the spherical symmetry of the Schwarzschild spacetime, one can decompose it as a warped product of two bi-dimensional geometries. In terms of the spacetime manifold, this can be written as: $\mathcal{M}_0 = M_2 \times_r S^2$, where $M_2$ has a Lorentzian geometry and $S^2$ is the 2-sphere. In terms of the metric line element we can write:
\begin{equation}
d\hat{s}{}^2 = g^{}_{ab}(x^c)dx^a dx^b + r^2(x^a) \Omega^{}_{AB}(x^C) dx^A dx^B\,,
\label{metric-schwarzschild-warped-product}
\end{equation}
where $g^{}_{ab}$ is the Lorentzian metric of $M_2$ and $\Omega_{AB}$ is the metric on the 2-sphere $S^2$. In the case of the Schwarzschild metric we can write
\begin{eqnarray}
g^{}_{ab} dx^a dx^b & = & -f(r)\,dt^2+\frac{dr^2}{f(r)} \,,
\label{metric-on-M2} \\
\Omega^{}_{AB} d\Theta^A d\Theta^B & = & d\theta^2 +\sin^2\theta\, d\varphi^2 \,, \label{metric-on-S2}
\end{eqnarray}
where $f(r)$ is given in Eq.~\eqref{fsch-expression} and we have chosen $x^{a}=(t,r)$ and $\Theta^{A}=(\theta, \varphi)$.
In Eq.~\eqref{metric-schwarzschild-warped-product}, $r(x^a)$ is the areal radial coordinate and $r^2$ is the geometry warp factor. The covariant derivative on $S^2$ is denoted by a vertical bar ($\Omega_{AB|C} = 0$) while for $M^2$ we use a colon ($g_{ab:c} = 0$). The volume Levi-Civita tensor in $S^2$ is denoted by $\epsilon_{AB}$ while in $M^{2}$ is denoted by $\varepsilon_{ab}\,$. 
Something interesting about the metric of $M^2$ [see Eq.~\eqref{metric-on-M2}] is that, if we use the {\em tortoise}  coordinate
\begin{equation}
\frac{dx}{dr} = \frac{1}{f}\,,    
\label{general-definition-tortoise-coordinate}
\end{equation}
the metric can be rewritten as:
\begin{eqnarray}
ds^2_{{M^2}} =  -f(r)\left( dt^2 + dx^2 \right) 
\,,
\label{metric-on-M2-conformally-flat} 
\end{eqnarray}
that is, in an explicitly conformally-flat form (remember that any metric of a two-dimensional manifold is locally conformally flat).  Here, we have to interpret $r$ as a function of $x$, such that [see Eq.~\eqref{general-definition-tortoise-coordinate}]
\begin{equation}
\frac{dr(x)}{dx} = f(r(x))\,.    
\end{equation}
It is important to stress that while the perturbative approach linearizes the dynamics of fast degrees of freedom, it provides a simplified but non-trivial framework to investigate the presence of (universal) integrable structure governing the slow/background DoFs. Indeed, while the dynamics\\
 of perturbations is linear by construction [see Eq.~\eqref{first-order-eqs}], the background is a fully nonlinear solution with a rich structure acting on the perturbations.

\subsection{Linear dynamics of the \emph{fast} DoFs: master equations}
\label{fast-dofs-master-eqs}

The spherically symmetric structure of the background suggests to expand the metric perturbations in tensor spherical harmonics in order to separate, in the Einstein perturbative equations, the dependence on the coordinates of $M^2$ from the angular dependence. Moreover, the tensor harmonics can be further separated into two classes according to how they transform under parity transformations [$(\theta,\phi)$ $\rightarrow$ $(\pi-\theta, \phi+\pi)$]. In fact, one can distinguish between odd parity tensors $\mathcal{O}^{\ell m}$, which transform as $\mathcal{O}^{\ell m} \rightarrow (-1)^{\ell+1}\mathcal{O}^{\ell m}$, and even parity tensors $\mathcal{E}^{\ell m}$ with the property of transforming as $\mathcal{E}^{\ell m} \rightarrow (-1)^{\ell}\mathcal{E}^{\ell m}$. 

The first step consists in splitting the metric perturbations into harmonics and parity, i.e. schematically
\begin{eqnarray}
h_{\mu\nu} = \sum_{\ell ,m} h^{\ell m, \rm odd}_{\mu\nu} + h^{\ell m, \rm even}_{\mu\nu} \,.
\label{metric-pert-harmonic-parity}
\end{eqnarray}
We are not going to show explicit expressions for the harmonic components of the metric perturbations since they are not relevant in this work. They can be found, for instance, in Refs.~\cite{Lenzi:2021njy, Martel:2005ir}. We also drop the harmonic indices from now on for the sake of clarity.
The separation in Eq.~\eqref{metric-pert-harmonic-parity} allows to decouple 
Einstein perturbative equations into an infinite set of equations for the different $(\ell,m)$ and parity modes (see, e.g.~\cite{Gerlach:1979rw, Gerlach:1980tx}).
Now, let us consider the Einstein perturbative equations for just a single mode, that is, for fixed harmonic numbers $(\ell,m)$ and parity. Moreover, we focus on the radiative modes, i.e. $\ell \geq  2$, neglecting, for the time being, the non-radiating modes $\ell = 0,1$~\cite{Zerilli:1970la} (see also~\cite{Martel:2005ir}). These equations, seven for the even-parity case and three for the odd-parity case, are $1+1$ linear PDEs (i.e. only with dependence on the coordinates of $M^2$) that can be combined in such a way that we can isolate (decouple) special gauge-invariant combinations of the metric perturbations (including their first-order derivatives), known as {\em master functions}. In this way, the perturbative Einstein equations for a single mode  reduce to wave-type equations with a potential, known as {\em master equations}. The concrete form of the master equations is
\begin{equation}
\left(-\frac{\partial^2}{\partial t^2} + \frac{\partial^2}{\partial x^2} - V^{\rm even/odd}_\ell  \right)\Psi^{\rm even/odd} = 0\,,
\label{master-wave-equation}
\end{equation}
where $x$ is the tortoise coordinate introduced in Eq.~\eqref{general-definition-tortoise-coordinate}.
In the wave equation~\eqref{master-wave-equation}, $\Psi^{\rm even/odd}(t,r)$ is the even/odd-parity master function; and $V^{\rm even/odd}_\ell(r)$ is the associated potential, which only depends on the harmonic number $\ell$ and the parity. For the cases where we have explicit expressions of the potentials, these are in terms of the areal radial coordinate $r$, but for the purposes of solving the master wave-type equations, we have to understand this dependence on $r$ as on a function of the radial coordinate, i.e. on $r=r(x)$.

The decoupling of the perturbative Einstein equations for a single mode (fixed parity and harmonic numbers $\ell$ and $m$), was first carried out in the seminal paper of Regge and Wheeler~\cite{Regge:1957td}, and later by Cunningham, Price, and Moncrief~\cite{Cunningham:1978cp,Cunningham:1979px,Cunningham:1980cp} and also by  Zerilli and Moncrief~\cite{Zerilli:1970la,Moncrief:1974vm}. There are a number of works that develop further BHPT in the non-rotating case and its applications to gravitational wave astronomy, astrophysics in general, and high-energy physics (some relevant works, although not an exhaustive list, are~\cite{Gerlach:1979rw,Gerlach:1980tx,Kokkotas:1999bd,Nollert:1999re,Andersson:2000mf,Sarbach:2001qq,Clarkson:2002jz,Sasaki:2003xr,Kokkotas:2003mh,Ferrari:2007dd,Cardoso:2012qm,Brito:2015oca,Berti:2018vdi,Barack:2018ylyv}).
Once the master equations are solved, we can reconstruct all the metric perturbations from the master functions (see, e.g.~\cite{Hopper:2010uv,Martel:2003jj,Spiers:2023mor,Lenzi:2024tgk}).

Recently, in Ref.~\cite{Lenzi:2021wpc}, a systematic approach to the decoupling of the perturbative Einstein equations showed that there are actually infinite ways to reduce the perturbative Einstein equations to an infinite set of master equations. This infinite hierarchy of master equations can be classified into two distinct branches: a \textit{standard branch} and a \textit{Darboux branch}~\cite{Lenzi:2021njy} (see also Ref.~\cite{Lenzi:2024tgk} for the extension to the presence of perturbative sources). In the so-called standard branch, the most general master function can be written as follows
\begin{eqnarray}
{}^{}_{\rm S}\Psi^{\rm odd/even} =
\left\{ \begin{array}{lc}
\mathcal{C}^{}_1 \Psi^{\rm CPM} + \mathcal{C}^{}_2\Psi^{\rm RW} &  \mbox{odd parity}\,, \\[2mm]
\mathcal{C}^{}_1 \Psi^{\rm ZM} + \mathcal{C}^{}_2\Psi^{\rm NE} &  \mbox{even parity}\,,
\end{array} \right.
\end{eqnarray}
where $\mathcal{C}^{}_1$ and $\mathcal{C}^{}_2$ are arbitrary constants. Therefore, for odd-parity perturbations, the master function is a linear combination of the Regge-Wheeler (RW)~\cite{Regge:1957td} and the Cunningham-Price-Moncrief (CPM)~\cite{Cunningham:1978cp,Cunningham:1979px,Cunningham:1980cp} master functions. 
In the case of even-parity perturbations, the most general master function is instead a linear combination of the  Zerilli-Moncrief (ZM)~\cite{Zerilli:1970la,Moncrief:1974vm} and another one (NE) which was first found in covariant form in Ref.~\cite{Lenzi:2021wpc}~\footnote{This master function already appeared in Ref.~\cite{Gleiser:1998rw}, in Schwarzschild coordinates and in the Regge-Wheeler gauge, as a better choice for computations in the context of the close limit approximation.}. We are not going to show here the explicit covariant expressions as they can be found in Refs.~\cite{Lenzi:2021wpc, Lenzi:2024tgk}.

For each parity, in the standard branch, there is just one potential, the Regge-Wheeler (odd-parity) and Zerilli (even-parity) potentials, i.e.
\begin{eqnarray}
{}^{}_{\rm S}V^{\rm odd/even}_\ell =
\left\{ \begin{array}{lc}
V^{\rm RW}_\ell &~~  \mbox{odd parity}\,, \\[2mm]
V^{\rm Z}_\ell  &~~  \mbox{even parity}\,,
\end{array} \right.
\label{potentials-standard-branch}
\,,
\end{eqnarray}
where the Regge-Wheeler potential reads
\begin{equation}
V_{\ell}^{\rm RW}(r) = \left(1-\frac{r^{}_s}{r}\right)  \left(\frac{\ell(\ell+1)}{r^{2}} - \frac{3r^{}_{s}}{r^{3}} \right)  \,,
\label{schwarzschild-omega-potential-odd-parity}
\end{equation}
while the Zerilli potential can be written as follows
\begin{widetext}
\begin{equation}
V^{\rm Z}_\ell(r) = \frac{f(r)}{\lambda^{2}(r)}\left[ \frac{(\ell-1)^{2}(\ell+2)^{2}}{r^{2}}\left( \ell(\ell+1) + \frac{3r^{}_{s}}{r} \right)  + \frac{9r^{2}_{s}}{r^{4}}\left( (\ell-1)(\ell+2) + \frac{r^{}_{s}}{r} \right) \right] \,,
\label{schwarzschild-omega-potential-even-parity}
\end{equation}
\end{widetext}
where
\begin{eqnarray}
\lambda(r) =  (\ell-1)(\ell+2) + \frac{3r^{}_{s}}{r}\,. 
\label{lambda-def-sch}
\end{eqnarray}
The second branch is new, and is called the Darboux branch due to the emerging symmetry in the space of master functions and equations (see Sec.~\ref{Ss:darboux-covariance}). The Darboux branch contains an infinite class of pairs (master function, potential). It turns out that the potential can be any solution of the following nonlinear second-order ordinary differential equation (ODE): 
\begin{equation}
\left(\frac{\delta{V}^{}_{,x}}{\delta{V}} \right)^{}_{,x}
+ 2 \left(\frac{V^{\rm RW/Z}_{\ell,x}}{\delta{V}} \right)^{}_{,x} - \delta{V} = 0 \,,
\label{potentials-darboux-branch}
\end{equation}
where $\delta{V} = {}_{\rm D}V^{\rm odd/even}_\ell - V^{\rm RW/Z}_\ell$. The master functions in the Darboux branch depend explicitly on the potential. Given a potential, the master functions read~\cite{Lenzi:2021wpc, Lenzi:2024tgk}
\begin{eqnarray}
{}^{}_{\rm D}\Psi^{\rm odd} & = & \mathcal{C}^{}_1\left( \Sigma^{\rm odd}\, \Psi^{\rm CPM} + \Phi^{\rm odd} \right)\,,
\label{odd-master-function-darboux}
\\[2mm]
{}^{}_{\rm D}\Psi^{\rm even} & = & \mathcal{C}^{}_1\left( \Sigma^{\rm even}\, \Psi^{\rm ZM} + \Phi^{\rm  even} \right) \,,
\label{even-master-function-darboux}
\end{eqnarray}
where $\Sigma^{\rm odd/even}(r)$ are functions containing integrals of the potential and $\mathcal{C}_1$ is an arbitrary constant. Here, $\Phi^{\rm odd/even}$ are combinations of metric perturbations and their first derivatives, but they are not by themselves master functions. Only their combination with $\Psi^{\rm CPM/ZM}$ in Eqs.~\eqref{odd-master-function-darboux} and~\eqref{even-master-function-darboux} are master functions~\cite{Lenzi:2021wpc}. Again, explicit expressions of all these functions can be found in Ref.~\cite{Lenzi:2021njy} (see also Ref.~\cite{Lenzi:2024tgk} for some typographical corrections).

Each pair $(\Psi, V)$ in the infinite set of potentials and master functions  described above is a valid (physically equivalent, see Sec.~\ref{Ss:darboux-covariance}) choice of “slow” and “fast” DoFs as seen from the perspective of point iii) in Sec.~\ref{s:wave_mean-flow_asymp-reason}: the potential $V$ is completely determined by the background and corresponds to the frozen slow DoFs while the master functions are the fast DoFs with linear dynamics.

 \subsection{Darboux covariance of BH perturbations}
\label{Ss:darboux-covariance}

Let us now provide a first hidden structure in the landscape of master equations and functions described above, which was found in Ref.~\cite{Lenzi:2021njy} and has been called it there \textit{Darboux covariance}. This provides a first hint towards unveiling the “bulk” underlying integrable structures in BHPT, which will be investigated in Sec.~\ref{Ss:bulk-cauchy}.

It was first noted by Chandrasekhar~\cite{Chandrasekhar:1975zza,Chandrasekhar:1979iz} (see also~\cite{Heading:1977jh,1980RSPSA.369..425C}) that odd- and even-parity sectors, while being sharply separated, share a deep relation. In order to show Chandrasekhar's relation between the two sector one has to consider single-frequency solutions, i.e.
\begin{equation}
\Psi(t,r) = e^{ik t}\,\psi(k;x) \,,
\label{time-to-frequency-domain}
\end{equation}
thus reducing the master equation to a time-independent Schr\"odinger equation
\begin{equation}
\mathcal{L}^{}_V \psi \equiv \psi^{}_{,xx} - V\psi = -k^2\psi \,.
\label{schrodinger}
\end{equation}
Starting from what is usually called in the literature an \textit{algebraically special} solution~\cite{Chandrasekhar:1984:10.2307/2397739}~\footnote{In a scattering context this is usually named \textit{antibound state}.}, i.e.
\begin{equation}
    \psi^{}_0 = \frac{\lambda(r)}{2}e^{-i k^{}_0 x}  \,,
    \label{algebraically-special}
\end{equation}
with
\begin{equation}
k^{}_0 = -i\frac{(\ell+2)(\ell+1)\ell(\ell-1)}{12\,M}\,,
\label{algebraically-special-frequency}
\end{equation}
Chandrasekhar was able to show that the Regge-Wheeler and Zerilli potentials are related by the following relation~\footnote{This expression, which is different from the (analogous) original one in Ref.~\cite{1980RSPSA.369..425C}, fits better the purposes of this work.}
\begin{equation}
V^{\mathrm{RW}}_{\mathrm{Z}} =  \mp W^{}_{0,x}  + W_{0}^2 + k_0^2\,, 
\label{RWZ-superpotential}
\end{equation}
where
\begin{eqnarray}
\nonumber
W^{}_{0}(r) &=& -(\ln{\psi^{}_0})^{}_{,x}
\\
&=&
i k^{}_0 + \frac{6\,M f(r)}{\lambda(r)r^2} \,.    \label{generating-function-RWZ}
\end{eqnarray}
The consequence of this relation between the two potentials is the isospectrality of the two parity sectors, namely they share the same physical description in frequency domain~\cite{Chandrasekhar:1975zza,1980RSPSA.369..425C}. It was only recently realized~\cite{Glampedakis:2017rar} that this physical correspondence is due to the presence of a Darboux transformation~\cite{Darboux:89, Matveev:1991ms}, which in frequency domain can be defined as the transformation in Eq.~\eqref{RWZ-superpotential}. As we can see in Eq.~\eqref{generating-function-RWZ}, this transformation is determined by the logarithmic derivative of a particular solution of the master wave equation.

In Ref.~\cite{Lenzi:2021njy}, the typical frequency domain definition of Darboux transformations~\cite{Darboux:89, Matveev:1991ms} was extended  to the full time domain master equation in order to reveal that all the pairs $(\Psi, V)$ are related by Darboux transformations. 
Starting from any two master equations in the following form
\begin{equation}
\left( -\partial^2_t + \mathcal{L}^{}_v \right) \Phi = 0\,,  \qquad
\left( -\partial^2_t + \mathcal{L}^{}_V \right) \Psi = 0 \,.  
\label{two-master-equation}
\end{equation}
a Darboux transformation is defined by the following system of equations 
\begin{eqnarray}
(\Phi, v) \rightarrow (\Psi, V):\quad \left\{ \begin{array}{lcl}
\Psi & = & \Phi^{}_{,x}  + W\,\Phi \,, \\[2mm]
V & = & v + 2\,W^{}_{,x}\,,
\end{array} \right.
\label{Darboux-transformation}
\end{eqnarray}  
where $W$ is the Darboux transformation generating function, which has to satisfy a consistency condition, in the form of a Riccati equation
\begin{eqnarray}
 W^{}_{,x} - W^2 + v = \mathcal{C}^{}_R \,,   
\label{riccati-g}
\end{eqnarray}
where $\mathcal{C}_R$ is an arbitrary integration constant. Notice that logarithmic derivatives of a solution to the time independent Schr\"odinger equation automatically satisfy the Riccati equation, e.g. $W_{0}$ satisfies it with $\mathcal{C}^{}_R=k_0^2$. The consistency between the expressions for $W$ and $W_{,x}$, as found from Eqs.~\eqref{Darboux-transformation} and~\eqref{riccati-g}, gives the following equation for the difference $\delta V = V-v\,$ 
\begin{equation}
\left(\frac{\delta V^{}_{,x}}{\delta V} \right)^{}_{,x}
+ 2 \left(\frac{v^{}_{,x}}{\delta V} \right)^{}_{,x} - \delta V = 0 
\,, 
\label{xdarboux}
\end{equation}
which coincides with Eq.~\eqref{potentials-darboux-branch}, i.e. the equation that every potential in the Darboux branch has to satisfy~\cite{Lenzi:2021wpc}, with the identification $v=V^{\rm RW/Z}$. The conclusion is that the infinite hierarchy of master equations obtained starting from the Einstein perturbative equations, is connected by Darboux transformations and thus provides infinite physically equivalent descriptions of the linear dynamics of gravitational perturbations around a spherically symmetric BH~\cite{Lenzi:2021njy} (see also~\cite{Lenzi:2024tgk} for more details). In this context, Chandrasekhar transformation~\eqref{RWZ-superpotential} correspond to the Darboux transformation connecting the two parity sectors. In the light of the program proposed in Sec.~\ref{s:intro}, we can say that Darboux covariance guarantees that the infinite pairs $(\Psi, V)$ are physically equivalent choices of the slow and fast DoFs.

Finally, it is important to mention that the Darboux generating function is actually a superpotential in the supersymetric (SUSY) meaning. Indeed, using Eqs.~\eqref{Darboux-transformation} and~\eqref{riccati-g} one can write the BH potentials as
\begin{eqnarray}
    v = -W^{}_{,x} + W^2 + \mathcal{C}_{\mathrm{R}}\,,\quad
    V = +W^{}_{,x} + W^2 + \mathcal{C}_{\mathrm{R}}\,.
\end{eqnarray}
One can then shift the potentials by the constant $\mathcal{C}_{\mathrm{R}}$, i.e. $ V_{+} \equiv V-\mathcal{C}_{\mathrm{R}}$ and $V_{-} \equiv  v-\mathcal{C}_{\mathrm{R}}$, so that
\begin{eqnarray}
V^{}_{+} =  + W^{}_{,x} + W^2 \,, \quad  
V^{}_{-} =  - W^{}_{,x} + W^2 \,,
\label{darboux-superpotential}
\end{eqnarray}
and consequently, the frequency can be shifted as $\hat{k}= k-\mathcal{C}_{\mathrm{R}}$ so that
\begin{eqnarray}
\mathcal{L}^{}_{V_{-}}\phi & = & \left(\partial^{2}_{x} -V^{}_{-}\right)\phi = -\hat{k}^2\phi \,,  \\[2mm]
\mathcal{L}^{}_{V_{+}}\psi & = & \left(\partial^{2}_{x} -V^{}_{+}\right)\psi = -\hat{k}^2\psi  \,.
\end{eqnarray}
This helps writing the Schr\"odinger operators $\mathcal{L}_{V_{\pm}}$ in the SUSY factorized form as follows
\begin{eqnarray}
\mathcal{L}^{}_{V^{}_{-}}&=& \left(\partial^{}_x - W \right) \left(\partial^{}_x + W \right)\,, \\[2mm]
\mathcal{L}^{}_{V^{}_{+}}&=& \left(\partial^{}_x + W \right) \left(\partial^{}_x - W \right)\,.
\end{eqnarray}
Notice that Eq.~\eqref{RWZ-superpotential} already shows the superpotential relation between the Regge-Wheeler and Zerilli potentials~\cite{Heading:1977jh,1980RSPSA.369..425C}. Then, the (master) Schr\"odinger equations
\begin{equation}
\mathcal{L}^{}_{V^{\rm Z}_{\rm RW}} \psi^{\rm Z}_{\rm RW} = -k^2\psi^{\rm Z}_{\rm RW}\,,
\end{equation}
can be rewritten as
\begin{equation}
\mathcal{L}^{}_{\hat{V}^{\rm Z}_{\rm RW}} \psi^{\rm Z}_{\rm RW} = -\hat{k}^2\psi^{\rm Z}_{\rm RW} \,,
\end{equation}
where now
\begin{equation}
\hat{V}^{\rm Z}_{\rm RW} = \pm W^{}_{0,x} + W_0^2 \,, \qquad
\hat{k}^2 = k^2 -k_0^2\,.
\end{equation}
To know more about Darboux transformations in the context of SUSY quantum mechanics see, for instance, Refs.~\cite{Cooper:1982dm, Cooper:1994eh, Cevik:2016mnr, Gomez-Ullate:2003wul,GomezUllate:2004}, where, in particular, one can see the different role of Darboux transformations depending on whether the states (wave master functions) used to generate them are bounded or not.

\section{Integrable structures in Cauchy slices}
\label{Ss:bulk-cauchy}

We have described the complete landscape of master equations describing BH perturbations, i.e. pairs of master functions and potentials. We have also seen the role of the Darboux transformations as a way of connecting all of them and establishing the isospectrality of the underlying physical description corresponding to this infinite set. In this sense, the Darboux transformation plays the role of a covariance or symmetry of this set of possible physical descriptions of (vacuum) BH perturbations, establishing their physical equivalence. These results provide the correct way to separate the slow and fast DoFs in this context. In this section, we show the hidden integrable structure associated to the nonlinear slow DoFs, i.e. the potentials.~\footnote{Note, however, that the integrable structures we identify do not consider the non-radiative $\ell =0,1$ modes, which deserve a separate (future) investigation, possibly involving the need of a refinement/renormalization in the appropriate separation of fast/slow DoFs (we thank the anonymous referee for stressing this point)}.

\subsection{Korteweg-de Vries isospectral symmetries}
\label{Sec:kdv-symmetries}

\subsubsection{Darboux transformation and integrability of the KdV equation}

The Darboux transformation offers, both formally and historically, a direct connection between the linear PDE system described by the time-independent Schr\"odinger equation~\eqref{schrodinger} and a third-order nonlinear (dispersive) PDE, namely the Korteweg-de Vries (KdV)\footnote{The authorship of the KdV has also been assigned to Boussinesq~\cite{Boussinesq:1871jb,Boussinesq:1872jt}. See~\cite{deJager:2006em} for a discussion on the origin of the KdV equation.} equation~\cite{Korteweg:10.1080/14786449508620739}:
\begin{equation}
{\rm KdV}[V] \equiv V^{}_{,\sigma} - 6 VV^{}_{,x} + V^{}_{,xxx} = 0 \,,
\label{kdv-equation}
\end{equation}
where $\sigma$ is the time of the nonlinear KdV dynamics\footnote{Notice that in previous works~\cite{Lenzi:2021njy,Lenzi:2022wjv,Lenzi:2023inn} we used $\tau$ for the KdV evolution instead of $\sigma$. Here we use $\tau$ for the time along the hyperboloidal slicing.}, which should not be confused with the Schwarzschild time $t$. From the point of view of BH physics, $\sigma$ does not play in principle any role. One interesting analogy with the original KdV equation, which was introduced to study the propagation of waves in shallow water, is that in this context $V$ is essentially the approximation to the gravitational Newtonian potential around the Earth surface, while in our case it is going to be exactly any of the BH potentials (see Sec.~\ref{fast-dofs-master-eqs}).

In the pioneering  work by Gardner, Greene, Kruskal and Miura~\cite{Gardner:1967wc}, it was shown that the KdV equation is exactly solvable by using the inverse scattering method~\cite{Deift:1979dt,Ablowitz:1981jq,Eckhaus:1981tn,Novikov:1984id}.
This means that we can express the potential $V$ in terms of the spectral and scattering data of the Schr\"odinger operator in Eq.~\eqref{schrodinger}. Exact solvability is often related to the presence of infinite constants of integration~\cite{Zakharov:1991xd}. It was first realized that KdV equation possess infinite conserved integrals~\cite{Miura:1968JMP.....9.1204M} by making use of a particular nonlinear transformation for the potential $V$, called the Miura transformation~\cite{Miura:1968JMP.....9.1202M}:
\begin{eqnarray}
    V = U^{}_{,x} + U^2 \,,
    \label{miura-transformation}
\end{eqnarray}
which maps solutions to the KdV equation into solutions of the so-called modified KdV (mKdV) equation. In other words, given a solution $V$ of the KdV equation then $U$, as defined by the Miura transformation of Eq.~\eqref{miura-transformation}, is a solution of the following mKdV equation:
\begin{eqnarray}
{\rm KdV}[V] & = & \left(2\,U \pm \partial^{}_{x}\right) \left( U^{}_{,\sigma} - 6U^2 U^{}_{,x} + U^{}_{,xxx} \right) \nonumber \\[2mm] 
& = & \left(2\,U \pm \partial^{}_{x}\right)  {\rm mKdV}[U] \,, 
\label{mKdV}
\end{eqnarray}
Notice that the Miura transformation~\eqref{miura-transformation} is actually equivalent to the SUSY form of the potential in Eq.~\eqref{darboux-superpotential}, with the plus sign, which in turn comes from the Darboux relation between two different potentials. The connection between the KdV equation and the Schr\"odinger equation becomes then apparent. The scaling with the Riccati constant $\mathcal{C}_{R}$ is irrelevant since the KdV equation possesses Galilean invariance~\cite{1791585}. This shows an intimate relation between the KdV equation and the Darboux transformations: starting from a solution $V(\sigma,x)$ of the KdV equation, one can always find a pair of isospectral potentials, related by a Darboux transformation, where the Darboux generating function is found by solving the Miura transformation/Riccati equation~\eqref{miura-transformation}. Therefore, we can say that any snapshot (fixed KdV time, say $\sigma_0$) of a KdV solution, $V(\sigma_0,x)$, generates, through a Riccati equation, a Darboux transformation between two associated linear systems with different, but isospectral, potentials defined, as in Eq.~\eqref{darboux-superpotential}, by $V_{\pm} = \pm U_{,x} +U^2$, where $V_{+} = V$. Notice, however, that we still have not yet investigated the spectral properties of the KdV equation, and hence we cannot make any statement regarding the spectral relation among these two Darboux related potentials and a different KdV-time snapshot, say $V(\sigma_1,x)$. All this motivated the study of KdV deformations of the master equation as a “higher level” integrable structure of the dynamics~\cite{Lenzi:2021njy,Lenzi:2022wjv,Lenzi:2023inn}.

\subsubsection{Isospectrality under KdV flows}
 
There are different points of view to study the spectral properties of the time-independent Schr\"odinger equation under deformations governed by the KdV equation. 
A particularly elegant and powerful approach is the formulation due to Lax~\cite{Lax:1968fm}, who found a way to associate nonlinear evolution operators to linear operators, in such a way that the eigenvalues of the linear operator are first integrals (conserved quantities) of the nonlinear evolution. Lax was motivated by the fact that this is exactly what happens when considering a KdV equation together with the time independent Schr\"odinger equation~\cite{Miura:1968JMP.....9.1204M}. Then, Lax realized that this result can be generalized to other systems by first finding the nonlinear operator associated to the linear one. This is the most tricky part of the procedure but can be solved with pseudo-differential operators~\cite{Babelon:2003qtg}. The nonlinear operator associated to the Schr\"odinger operator $\mathcal{L}_V$ is found to be
\begin{eqnarray}
P^{}_V = -4 \partial^{3}_{x} + 6 V\partial^{}_x + 3 V^{}_{,x}  \,.
\end{eqnarray}
The operators $(P^{}_V,\mathcal{L}^{}_V)$ form a Lax pair, that is, a pair of operators whose commutator is the action of an operator corresponding to an integrable non-linear PDE, the KdV equation in this case, times the unity operator:
\begin{equation}
\frac{\partial \mathcal{L}^{}_V}{\partial\sigma} - \left[P^{}_V,\mathcal{L}^{}_V\right] = -{\rm KdV}[V]\cdot {\rm Id} \,.
\label{Lax-pair-equation}
\end{equation}
Assuming that $V$ satisfies the KdV equation~\eqref{kdv-equation}, Eq.~\eqref{Lax-pair-equation} guarantees that the operator $\mathcal{L}_V$, which then depends on $\sigma$, remains unitarily equivalent, meaning that there is a one parameter family of unitary operators $A(\sigma)$ such that
\begin{equation}
\mathcal{L}^{}_V(\sigma) = A(\sigma) \cdot \mathcal{L}^{}_V(0) \cdot A^{-1}(\sigma)\,,
\end{equation}
where the unitary operator $A(\sigma)$ evolves as:
\begin{equation}
\frac{dA(\sigma)}{d\sigma} = P^{}_V\cdot A(\sigma) \,.
\end{equation}
This implies that the spectral properties of the operator $\mathcal{L}_V$ are preserved by the KdV flow, i.e. the evolution along the KdV-time $\sigma$. This establishes a connection between certain integrable systems and isospectral problems. 

Actually, if we consider a deformation of our time-independent Schr\"odinger equation~\eqref{schrodinger} as follows
\begin{equation}
\label{kdv-deformation}
\left\{
\begin{array}{l}
V(x)\rightarrow V(\sigma,x)
\\
\psi(x)\rightarrow\psi(\sigma,x)
\\
k\rightarrow k(\sigma)
\end{array}
\right.
\end{equation}
such that the wave function $\psi(\sigma,x)$ evolves according to the operator $P_V$:
\begin{eqnarray}
\psi^{}_{,\sigma} = P^{}_V \psi \,,
\label{Eq2-KdV-like-2}
\end{eqnarray}
we have~\cite{Lenzi:2021njy}
\begin{equation}
\left( {\rm KdV}[V]  - (k^{2})^{}_{,\sigma} \right)\psi = 0 \,.
\label{KdV-and-eigenvalue-deformation}
\end{equation}
This means that if the potential (the Schr\"odinger equation) is deformed according to the KdV equation we must have
\begin{equation}
\psi\,(k^{2})^{}_{,\sigma} = 0\,,
\end{equation}
and therefore, the square of the frequency, $k^2$, is invariant under the KdV flow.

We can carry this whole picture to the context of BHPT. Then, if we deform the BH potential as in Eq.~\eqref{kdv-deformation}, we immediately deduce that the whole spectrum is preserved by the KdV evolution. In particular, this is  also true for the QNM complex frequencies, as the whole derivation does not require any integrability condition on the wave function $\psi$. 

Let us end this subsection with some remarks on the relation between the KdV flows and the Darboux transformation. We have seen that the KdV deformation, actually any deformation that follows any of the infinite hierarchy of KdV equations, of the time-independent Schr\"odinger equation~\eqref{schrodinger} constitutes an isospectral symmetry of the dynamics of BH perturbations. In addition, there is a strong connection between this set of deformations and the Darboux transformations. Following the work of~\cite{Matveev:1991ms} (see also~\cite{Lenzi:2021njy,Lenzi:2022wjv}), it is possible to show that the system of equations
\begin{equation}
\mathcal{L}^{}_V \psi = -k^2\psi\,,\qquad
\partial^{}_\sigma\psi = P^{}_V\psi\,, 
\label{schrodinger-deformed}
\end{equation}
is invariant under Darboux transformations provided the Darboux transformation generating function, $W$, is KdV-deformed as 
\begin{eqnarray}
W^{}_{,\sigma} = -W^{}_{,xxx} + 6(V+W^{}_{,x})W^{}_{,x} \,.
\label{kdv-deformation-g}
\end{eqnarray}
Notice that this equation can actually be recast, using the Darboux relation~\eqref{Darboux-transformation}, as a mKdV equation~\eqref{mKdV}. This provides the inverse relation between KdV deformations and Darboux transformations: given a Darboux transformation whose generating function satisfies a mKdV equation, then this automatically defines, through the Miura transformation, the KdV deformation of the two Darboux related potentials.
\subsubsection{(Bi-)Hamiltonian formulation of KdV equation}

The KdV equation, as well as other integrable systems, admits a particular bi-Hamiltonian structure, meaning that one can find two different (but compatible) Poisson structures and Hamiltonians such that the KdV equation can be described in terms of two different by analogous Hamiltonian flows~\cite{Babelon:2003qtg}. Starting from the existence of infinite conservation laws~\cite{Miura:1968JMP.....9.1204M}, Gardner was the first to realize that the KdV equation can be cast as a Hamiltonian system~\cite{gardner1971korteweg}. A different approach was taken by Zakharov and Faddeev~\cite{Zakharov:1971faa}, who combined inverse scattering techniques together with the Hamiltonian formalism to find the action-angle variables of the system in terms of the Bogoliubov coefficients, which completely defines the scattering matrix in the linear Schr\"odinger system.

The presence of infinite conservation laws~\cite{Miura:1968JMP.....9.1204M} is actually deeply related to the fact that each conserved quantity of the KdV equation defines a higher order KdV flow:
\begin{eqnarray}
\partial^{}_{\sigma^{}_k} V = \mathcal{K}^{}_k[V]=\mathcal{D} \frac{\delta \mathcal{H}^{}_{k+1}[V]}{\delta V(x)} \,, \quad k=0,1,2,\ldots
\end{eqnarray}
where the $k=0$ case is a trivial one and $\mathcal{D} = \partial/\partial_x$.
The last equality shows that these evolution equations can actually be written as a Hamilton equation with Hamiltonian $\mathcal{H}^{}_{k+1}[V]$ ($k=0,1,2,\ldots$). These Hamiltonians are functionals of the potential $V$
\begin{equation}
\mathcal{H}^{}_k[V] = \int^{\infty}_{-\infty} dx\, h^{}_k(V,V^{}_{,x},V^{}_{,xx},\ldots) \,,
\label{kdv-integrals-h}
\end{equation}
which can be found through the recurrence relations due to Lenard~\cite{Praught:2005SIGMA...1..005P} (see also~\cite{gardner1971korteweg,Lax:1968fm}), which reads as follows
\begin{eqnarray}
\frac{\partial}{\partial x} \frac{\delta \mathcal{H}^{}_{k+1}}{\delta V} = 
\left(-\frac{\partial^3}{\partial x^3} +4V\frac{\partial}{\partial x} + 2 V^{}_{,x}\right) \frac{\delta \mathcal{H}^{}_{k}}{\delta V} \,.
\end{eqnarray}
Therefore, each (higher-order) KdV equation can be cast as a Hamiltonian system. After defining the typical Poisson brackets (see Appendix~\ref{Hamiltonian-Formalism}), one can write these equations as
\begin{equation}
\mathcal{K}^{}_k[V] = \left\{ V, \mathcal{H}^{}_{k+1}[V] \right\}^{}_{\rm GZF}\,.
\label{kdv-hierarchy-gzf}
\end{equation}
We can find the expressions for all the functionals $\mathcal{H}^{}_{k}[V]$ by just fixing the first density, for instance as follows
\begin{equation}
h^{}_0 = \frac{1}{2}V\quad \Rightarrow\quad \mathcal{H}^{}_0 = \frac{1}{2}\int^{\infty}_{-\infty} dx\, V(x) \,,
\end{equation}
where $\mathcal{H}^{}_0$ is usually known as the {\em mean height}, which is reminiscent of the motivation for the original derivation of the KdV equation discussed above. Then, from the Lenard recurrence, we get the next two conserved quantities, which are associated to the momentum of the KdV wave
\begin{equation}
h^{}_1 = \frac{1}{2}V^2\quad \Rightarrow\quad \mathcal{H}^{}_1 = \frac{1}{2}\int^{\infty}_{-\infty} dx\, V^2(x) \,,
\end{equation}
and to the energy of the wave
\begin{equation}
h^{}_2 = V^3 + \frac{1}{2}V^2_{,x}~\Rightarrow~\mathcal{H}^{}_2 = \int^{\infty}_{-\infty}\!\! dx \left(V^3 + \frac{1}{2}V^2_{,x}\right) \,,
\end{equation}
It is not difficult to show explicitly that these are conserved quantities by taking their evolution with respect to the KdV time $\sigma$. For this, one only has to consider the KdV equation and use the fact that the integral of total derivatives vanishes thanks to the decay of $V$ at $x\rightarrow\pm\infty$. 

As it has already been mentioned, the KdV equation possesses a second Hamiltonian structure, meaning that one can find a second Poisson bracket definition: the Magri brackets~\cite{Magri:1977gn} (see Appendix~\ref{Hamiltonian-Formalism}). Indeed, this additional Hamiltonian structure describes the same KdV Hamiltonian flows:
\begin{equation}
\mathcal{K}^{}_k[V] = \left\{ V, \mathcal{H}^{}_{k}[V] \right\}^{}_{\rm M} = \mathcal{E}^{}_V \frac{\delta \mathcal{H}^{}_{k}[V]}{\delta V(x)} \,,
\label{kdv-hierarchy-m}
\end{equation}
where the operator $\mathcal{E}^{}_V$ is defined as follows
\begin{equation}
\mathcal{E}^{}_V =  -\frac{\partial^3}{\partial x^3} +4V(x)\frac{\partial}{\partial x} + 2 V^{}_{,x}(x) \,.
\label{magri-operator}
\end{equation}
The KdV equation itself can be written as:
\begin{eqnarray}
\partial^{}_\sigma V = \mathcal{D}\frac{\delta \mathcal{H}^{}_{2}}{\delta V} 
= \mathcal{E}^{}_V \frac{\delta \mathcal{H}^{}_{1}}{\delta V} 
= 6VV^{}_{,x} - V^{}_{,xxx} \,.
\end{eqnarray}
Then, we can rewrite the Lennard recurrence in terms of the two operators associated to the two different Poisson brackets, i.e.
\begin{eqnarray}
\mathcal{D}\frac{\delta \mathcal{H}^{}_{k+1}[V]}{\delta V(x)} = \mathcal{E}^{}_V \frac{\delta \mathcal{H}^{}_{k}[V]}{\delta V(x)}  \,.
\end{eqnarray}
In summary, the KdV dynamics admits two Hamiltonian structures and, under both of them, we have an infinite number of conserved quantities, which we have called $\mathcal{H}^{}_k[V]$ ($k=0,1,2,\ldots$), and are functionals made out of differential polynomials in the potential $V$. From the different equations, and in particular using the Lenard recursion, we can see that these conserved quantities are in involution, that is, they satisfy:
\begin{equation}
\left\{  \mathcal{H}^{}_j ,  \mathcal{H}^{}_k \right\}^{}_{\rm GZF} =
\left\{  \mathcal{H}^{}_j ,  \mathcal{H}^{}_k \right\}^{}_{\rm M} = 0\,, \quad j,k = 0,1,2,\ldots
\end{equation}
therefore, we can say that the KdV equation is an integrable system in the classical Liouville sense~\cite{Arnold:1989am}.

\subsection{The structural triangle KdV-Virasoro-Schwarzian derivative}
\label{kdv-triangle}

In this section, we show how the structure of the KdV isospectral flows is intimately related to the Virasoro algebra of a conformal field theory (CFT) in two dimensions ($2d$). In doing so, we isolate the Schwarzian derivative~\eqref{schwarztian} as the mathematical object alerting for the presence of underlying conformal structures in the dynamics of perturbed BHs. This will be explored in Sec.~\eqref{Ss:bulk-cauchy}. Let us first consider the KdV equation as a bi-Hamiltonian system, as shown in Sec.~\ref{Sec:kdv-symmetries} and write it as a Hamiltonian flow:
\begin{eqnarray}
   \partial_{\sigma} V = \left\{ V, \mathcal{H}_2\right\}_{\mathrm{GFZ}} =
   \left\{ V, \mathcal{H}_1\right\}_{\mathrm{M}}
   \,.
   \label{kdv-hamiltonian-flow}
\end{eqnarray}
The interesting fact is that the Magri bracket $\left\{ \cdot , \cdot \right\}_{\mathrm{M}}$ [see Eq.~\eqref{Magri-bracket}], which is at the core of the second Hamiltonian formulation in Eq.~\eqref{kdv-hamiltonian-flow}, is actually deeply related to the (classical) Virasoro algebra~\cite{Gervais:1981gs,Gervais:1982nw,Gervais:1985fc,GERVAIS1985279, DiFrancesco:1997nk}. The Magri bracket of two potential functions can be evaluated as follows:
\begin{widetext}
\begin{eqnarray}
\nonumber
\left\{ V(x), V(y)\right\}^{}_{\mathrm{M}} & = & \int_{-\infty}^{\infty} dz \, \frac{\delta V(x)}{\delta V(z)}  \mathcal{E}^{}_V \frac{\delta V(y)}{\delta V(z)} 
= \int_{-\infty}^{\infty} dz\,  \delta(x-z)\left[-\partial_z^{3} + 4 V(z)\partial^{}_z + 2 V'(z)\right]\delta(y-z)  \\[2mm] \nonumber
& = & \partial_y^{3} \delta(y-x) - 4\,V(y)\, \partial^{}_y \delta(y-x) + 2 \,V'(y)\, \delta(y-x) \\[2mm]
& = & \partial_x^{3} \delta(x-y) - 4\,V(x)\, \partial^{}_x \delta(x-y) - 2 \,V'(x)\, \delta(x-y) \,,
\label{magri-VV}
\end{eqnarray}    
\end{widetext}
where the equalities in the second and third lines have to be considered in the distributional sense. 

The potential $V$ defines the generators of the Virasoro algebra. This can be seen by complexifying the coordinate $x$ ($x\rightarrow z\in \mathbb{C}$) and write the Laurent expansion of the potential $V$ as a function of the complex variable $z \in\mathbb{C}$, i.e.:
\begin{eqnarray}
    V(z) = \sum_{n=-\infty}^{\infty} L_n z^{-n-1}
    \,,
\end{eqnarray}
where the coefficients $L_n$ are defined as the following integrals over the unit circle ($|z|=1$):
\begin{equation}
    L^{}_n = \frac{1}{2\pi i} \oint dz\, V(z) z^{n+1} \,.
    \label{Virasoro-generators}
\end{equation}
Taking into account the action of the Dirac delta function in the complex domain:
\begin{equation}
    \oint dz\, \delta(z)f(z) = f(0) \,,
\end{equation}
the Magri brackets between the coefficients $\{L_n\}$ can be calculated starting from Eq.~\eqref{magri-VV}:
\begin{widetext}
\begin{eqnarray}
\left\{ L^{}_n, L^{}_m \right\}^{}_{\mathrm{M}} & = & \left(\frac{1}{2\pi i}\right)^2 \oint dz\, dw\,  z^{n+1} w^{m+1}\left\{ V(z), V(w)\right\}^{}_{\mathrm{M}} 
  \nonumber \\[2mm]
& = & \left(\frac{1}{2\pi i}\right)^2 \oint dz\, dw\, z^{n+1} w^{m+1} \left[ \partial_z^{3} \delta(z-w) -4\,V(z)\, \partial^{}_z \delta(z-w) - 2 \,V'(z)\, \delta(z-w)\right]    
   \nonumber \\[2mm]
& = &  \frac{1}{2\pi i} \left[-n(n^2 -1)\delta_{n+m,0} + 2 (n-m) L^{}_{n+m}  \right]
    \,,
\end{eqnarray}   
\end{widetext}
where, in the last line, we have used integration by parts. Therefore, the Magri bracket, intimately connected with the second Hamiltonian structure of the KdV equation, provides the classical realization of the Virasoro algebra for the generators $L_n$ with central charge $c=-6$, that is:
\begin{equation}
\pi i \left\{ L^{}_n, L^{}_m\right\}^{}_{\mathrm{M}} = (n-m)L^{}_{n+m} - \frac{n(n^2-1)}{2}\delta^{}_{n+m,0} \,.
    \label{virasoro-algebra}
\end{equation}
The KdV equation can then be related to a $2d$ CFT by the identification of the potential $V(z)$ with the holomorphic component, $T_{zz} = T(z)$, of the energy-momentum tensor of a $2d$ CFT\footnote{Given the connection with CFT, here we are using the notation of Ref.~\cite{DiFrancesco:1997nk}.}. One can in principle extend even more the analogy and also identify  the anti-holomorphic component of a 2d CFT energy-momentum tensor, $\bar{T}_{\bar{z}\bar{z}} =\bar{T}(\bar{z})$, with the complex conjugate of the potential $\bar{V}(\bar{z})$. Then, we can introduce the generators $\bar{L}_n$, which satisfy the same Virasoro algebra of Eq.~\eqref{virasoro-algebra} and the crossed Poisson/Magri bracket vanishes, i.e.:
\begin{eqnarray}
\pi i \left\{ \bar{L}^{}_n, \bar{L}^{}_m\right\}^{}_{\mathrm{M}} & = & (n-m)\bar{L}^{}_{n+m} - \frac{n(n^2-1)}{2} \delta^{}_{n+m,0} \,, \qquad
\\[2mm] 
\left\{ L^{}_n, \bar{L}^{}_m \right\}^{}_{\mathrm{M}} & = & 0 \,.
\end{eqnarray}
In two dimensions, any holomorphic mapping of the complex plane into itself represents a conformal transformation~(see, e.g.~\cite{DiFrancesco:1997nk}). We can write such a mapping as
\begin{equation}
    z\in \mathbb{C} \rightarrow w(z) \,.
\end{equation}
Then, the infinitesimal conformal transformation reads:
\begin{equation}
    w(z) = z + \epsilon(z)\,,
    \label{conformal-epsilon}
\end{equation}
Following Ref.~\cite{DiFrancesco:1997nk}, and keeping in mind the identification  between BH potential and the $2d$ CFT energy-momentum tensor, the infinitesimal conformal transformation of $V(z)$ is given by
\begin{equation}
\delta^{}_{\epsilon} V(w) = \frac{1}{2}\epsilon'''(w) - 2 V(w) \epsilon'(w) - \epsilon(w) V'(w) \,.
    \label{conformal-infinitesimal}
\end{equation}
Notice that this can actually be written in terms of the second KdV Hamiltonian as follows:
\begin{equation}
  \delta^{}_{\epsilon} V(w) = -\frac{1}{2}\mathcal{E}^{}_{\mathrm{V}}\,\epsilon(w)   = \left\{V(w),   F^{}_{\epsilon}\right\}^{}_{\mathrm{M}} \,,
  \label{kdv-conformal-flow}    
\end{equation}
where we have introduced the conformal charge
\begin{eqnarray}
    F^{}_{\epsilon} =  -\frac{1}{2} \int dz \,\epsilon(z) V(z) \,.
\end{eqnarray}
Finally, the exponentiation of the infinitesimal transformation~\eqref{conformal-infinitesimal} gives:
\begin{eqnarray}
    V(w) = \left(\frac{dw}{dz} \right)^{-2}\left[ V(z) +\frac{1}{2}\mathcal{S}(w(z)) \right]
    \,.
    \label{potential-schwarzian}
\end{eqnarray}
This can be checked a posteriori by taking its infinitesimal form. Here, $\mathcal{S}$ denotes the Schwarzian derivative:
\begin{equation}
\mathcal{S}\left(w(z)\right) \equiv \frac{ w^{}_{zzz} }{ w^{}_{z} } - \frac{3}{2}\left( \frac{w^{}_{zz}}{w^{}_{z}} \right)^2 \,.
\label{schwarztian}
\end{equation}
Therefore, the Magri operator $\mathcal{E}_{\mathrm{V}}$/Magri-bracket provides the infinitesimal conformal transformation of the KdV potential whose finite form can be written, as in Eq.~\eqref{potential-schwarzian}, in terms of the Schwarzian derivative of the conformal mapping.

This whole treatment is meant to settle the KdV symmetry within the BHPT framework (in Cauchy slices) as the integrable structure associated to the slow DoFs, which is the first step in the whole program described in the Introduction. Notice that the KdV symmetry seems to have a twofold structure which can be resumed with Eqs.~\eqref{kdv-hamiltonian-flow} and~\eqref{kdv-conformal-flow}, i.e.
\begin{eqnarray}
  \partial^{}_{\sigma} V  & = & 
   \left\{ V, \mathcal{H}^{}_1\right\}^{}_{\mathrm{M}} \,,
   \\[2mm]
   \delta^{}_{\epsilon} V & = & \left\{V, F^{}_{\epsilon}\right\}^{}_{\mathrm{M}} \,.
\end{eqnarray}
We are rewriting them together to highlight the fact that the fundamental structure responsible for both transformations (the KdV deformations and the conformal transformations) are the same and are given by the Magri bracket. 
Indeed, this is no coincidence as we can consider the KdV deformations as a restricted family of infinitesimal conformal transformations, as in Eq.~\eqref{conformal-infinitesimal}, by taking  
\begin{eqnarray}
    \epsilon(z) = V(z) \,.
\end{eqnarray}
With this choice, the infinitesimal conformal transformation of Eq.~\eqref{conformal-infinitesimal} becomes
\begin{equation}
\left.\delta^{}_{\epsilon}V(w)\right|^{}_{\epsilon=V} = 
\frac{1}{2}V'''(w) - 3 V(w) V'(w)  \,.
\end{equation}
This is the nonlinear dispersive part of the KdV equation~\eqref{kdv-equation}, which can then be rewritten as
\begin{eqnarray}
   \partial^{}_{\sigma} V = 2 \left.\delta^{}_{\epsilon}V\right|^{}_{\epsilon=V}  
   \,.
\end{eqnarray}
This clarifies the reason why the Magri bracket provides both infinitesimal conformal transformations and KdV deformations, since they are essentially the same structure.
We have already described in Sec.~\ref{Sec:kdv-symmetries} the physical consequences on BH perturbations of the KdV deformations, i.e. isospectrality and conserved quantities (the KdV integrals). In the next section, we are going to show how the KdV/Virasoro conformal structure leaks into BHPT. The Schwarzian derivative, being related to the exponentiation of the Magri brackets, gives the finite structure we are going to look for in the master equations for the BH perturbations.

\subsection{Conformal transformations in the frequency domain}
\label{Sec:conformal-cauchy}

In order to see the effects of the hidden conformal symmetry described in the previous section on the BH response to (gravitational) perturbations, it is convenient to consider the following general transformation on the time-independent Schr\"odinger equation~\eqref{schrodinger} 
\begin{eqnarray}
\left\{
\begin{array}{lcl}
x & \mapsto & x = x(\tilde{x}) \,, \\[2mm]
\psi & \mapsto & \psi(x) = \omega(\tilde{x})\tilde{\psi}(\tilde{x}) \,.
\end{array}
\right. \label{conformal-transformation-psi-x}
\end{eqnarray}
Using the following differential relations
\begin{eqnarray}
\partial^{}_{x} &  = & \tilde{x}'\partial^{}_{\tilde{x}}\,,  \qquad (\tilde{x}'\equiv d\tilde{x}/dx) \,, \\[2mm]   
\partial^{2}_{x} & = & \tilde{x}'' \partial^{}_{\tilde{x}} + (\tilde{x}')^2 \partial^2_{\tilde{x}} \,,
\end{eqnarray}
we have that the time-independent Schr\"odinger equation [see Eq.~\eqref{schrodinger}] can be written as follows:
\begin{widetext}
\begin{equation}
\omega (\tilde{x}')^2 \tilde{\psi}^{}_{,\tilde{x}\tilde{x}} + 
\left( \omega\tilde{x}'' + 2\omega^{}_{,\tilde{x}} (\tilde{x}')^2 \right)\tilde{\psi}^{}_{,\tilde{x}} +
\left( \tilde{x}''\omega^{}_{,\tilde{x}} + (\tilde{x}')^2 \omega^{}_{,\tilde{x}\tilde{x}} 
- \omega\,V\right) \tilde{\psi} = -k^2\omega\,\tilde{\psi} \,.
\label{newschro}
\end{equation}
\end{widetext}
We are going to restrict ourselves to transformations that cancel the linear term in $\tilde{\psi}_{,\tilde{x}}$, so that the final equation resembles the original Schr\"odinger equation [Eq.~\eqref{schrodinger}].  
By looking into Eq.~\eqref{newschro}, we see that cancelling the term in $\tilde{\psi}^{}_{,\tilde{x}}$ means to impose the following relation between $\omega$ and $\tilde{x}(x)$:
\begin{eqnarray}
\frac{\omega^{}_{,\tilde{x}}}{\omega} = -\frac{\tilde{x}''}{2(\tilde{x}')^2} = \frac{1}{2}\left(\frac{1}{\tilde{x}'} \right)' \,,  
\end{eqnarray}
or equivalently
\begin{equation}
\frac{\omega'}{\omega} = -\frac{1}{2}\frac{\tilde{x}''}{\tilde{x}'}\,,
\end{equation}
which is a simple ODE that can be solved for $\omega$. The result is:
\begin{eqnarray}
\omega = \frac{\omega^{}_o}{\sqrt{\tilde{x}'}}  =  \omega^{}_o \sqrt{x^{}_{\tilde{x}}} \,, \label{omega-expression-cauchy}
\end{eqnarray}
where $\omega_o$ is an integration constant. Then, using this expression for $\omega$ and taking into account the following differential relations
\begin{eqnarray}
\tilde{x}' & = & \frac{1}{x^{}_{\tilde{x}}}  \,, \\[2mm]
\tilde{x}'' & = & - \frac{x^{}_{\tilde{x}\tilde{x}}}{x^{3}_{\tilde{x}}} \,, \\[2mm]
\tilde{x}''' & = & \frac{1}{x^{4}_{\tilde{x}}} \left( - x^{}_{\tilde{x}\tilde{x}\tilde{x}} + 3 \frac{x^{2}_{\tilde{x}\tilde{x}}}{x^{}_{\tilde{x}}}  \right) \,,
\end{eqnarray}
we can finally write the \emph{new} time-independent Schr\"odinger equation [Eq.~\eqref{newschro}] as:
\begin{equation}
\tilde{\psi}^{}_{\tilde{x}\tilde{x}} + \left( k^2 - V \right) x^{2}_{\tilde{x}}\, \tilde{\psi} + \frac{1}{2}\mathcal{S}\left(x(\tilde{x})\right) \tilde{\psi} = 0 \,, 
\label{new-sch}
\end{equation}
where $\mathcal{S}$ denotes the Schwarzian derivative defined in Eq.~\eqref{schwarztian}. An interesting property of the Schwarzian derivative is:
\begin{equation}
\mathcal{S}\left(\tilde{x}(x)\right) = -\frac{1}{x^2_{\tilde{x}}} \mathcal{S}\left(x(\tilde{x})\right) =
-\left(\tilde{x}'\right)^2 \mathcal{S}\left(x(\tilde{x})\right) \,.
\label{property-schwarzian}
\end{equation}

Finally, we can rewrite Eq.~\eqref{new-sch} in a way that looks almost identical to the time-independent Schr\"odinger equation:
\begin{equation}
\tilde{\psi}^{}_{\tilde{x}\tilde{x}} + \left( k^2 x^{2}_{\tilde{x}} - \tilde{V}  \right)\tilde{\psi} = 0 \,,  
\label{new-sch-2}
\end{equation}
where [using Eq.~\eqref{property-schwarzian}]
\begin{eqnarray}
\tilde{V} & = & x^{2}_{\tilde{x}}\,V - \frac{1}{2} \mathcal{S}\left(x(\tilde{x})\right) = \nonumber \\[2mm]
& = & x^{2}_{\tilde{x}}\left( V + \frac{1}{2} \mathcal{S}\left(\tilde{x}(x)\right) \right)\,.
\end{eqnarray}
Therefore, the potential transforms according to the standard conformal transformation of (the holomorphic components of) the energy-momentum tensor of a $2d$ CFT, i.e.
\begin{eqnarray}
V(x) \mapsto \tilde{V}(\tilde{x}) = \left(\frac{d \tilde{x}}{d x}\right)^{-2}\left(V(x) + \frac{1}{2}\mathcal{S}(\tilde{x}(x))\right)
\,, \quad
\end{eqnarray}
as in Eq.~\eqref{potential-schwarzian}.  This means that the potential transforms as a “quasi-primary” field, where the Schwarzian derivative measures the deviation from a primary field behaviour, which is recovered if and only 
if the Schwarzian derivative vanishes, $\mathcal{S}(\tilde{x}(x)) = 0$,
which in turn characterizes $x=x(\tilde{x})$ as a M\"obius
transformation. 

Let us now make some useful remarks regarding this construction, although we do not fully address them in this work as they require a significant number of new developments.
First, notice that in Eq.~\eqref{new-sch-2} the eigenvalue $k^2$ appears now multiplied by a factor $x^{2}_{\tilde{x}}\,$. In future investigations, it would be interesting to see whether a full conformal transformation of the Lorentzian manifold $M_2$, i.e. of the time domain equation, can account for such a multiplicative factor. Then, under the choice in \eqref{omega-expression-cauchy} for $\omega(\tilde{x})$, 
and choosing $\omega_o = 1$ since the equations are independent from it, the inverse of the transformation \eqref{conformal-transformation-psi-x} for $\psi(x)$
reads
\begin{eqnarray}
\left\{
\begin{array}{lcl}
x & \mapsto & \tilde{x} = \tilde{x}(x) \,, \\[2mm]
\psi(x) & \mapsto & \tilde{\psi}(\tilde{x}) = (\omega(x))^{-1}\psi(x) \\
& & = (\sqrt{x_{\tilde{x}}})^{-1}\psi(x) = \left(\frac{d \tilde{x}}{d x}\right)^{1/2} \psi(x) \,.
\end{array}
\right. \label{wirggconformal-transformation-psi-x}
\end{eqnarray}
Notice that the transformation of the wave function $\psi$ reminds the conformal transformation, under 
the mapping $x\mapsto \tilde{x}=\tilde{x}(x)$, of a primary field~\cite{DiFrancesco:1997nk}. Indeed, the KdV/Virasoro structure described in Sec.~\ref{kdv-triangle} relies on the extension to the complex plane (complexification of the independent variable $x$) and on treating the new complex variables $z$ and $\bar{z}$ as independent. The conformal transformation of a primary field in this space is then written as
\begin{eqnarray}
\bar{\Phi}(w,\bar{w}) = \left(\frac{d w}{d z}\right)^{-\Delta} \left(\frac{d \bar{w}}{d \bar{z}}\right)^{-\bar{\Delta}} \Phi(w,\bar{w})  \,.
\end{eqnarray}
After identifying $(x, \tilde{x})$ with $(z,w)$, we realize that the above inverse transformation on $\psi$ is just the holomorphic part of the conformal transformation of a primary field in two dimensions, with the primary field having holomorphic conformal dimension $\Delta = -1/2$.

For static states (equivalent to the zero frequency limit), similar transformations have been associated to the so-called Schr\"odinger symmetry and to the vanishing of tidal Love numbers~\cite{BenAchour:2022uqo}, i.e. constants related to (non) tidal deformability of BHs (and other compact bodies). A deeper understanding of the symmetry structure related to $\psi$ in the non-zero frequency limit (and also in time domain) is outside the scope of the present paper and will be postponed to future investigations. Indeed, what defines the slow/bulk DoFs is actually the solitonic potential characterizing the BH response to perturbations and the proper analysis of the wave functions will inevitably mix with boundary/asymptotic information. 

Both the transformation of the wave function, reminiscent of a conformal transformation for a primary field, and the multiplicative factor in front of the frequency term, suggest that we may have to consider a full conformal transformation of the spacetime (or at least of the two dimensional Lorentzian manifold) in order to gain more insight into these features.

%
\section{Integrable structures in hyperboloidal slices}
\label{Ss:bulk-hyperboloidal}

We have investigated the emergence of an underlying integrable structure associated to the slow DoFs, i.e. the solitonic potential response of the BH to gravitational perturbations, appearing when studying the linear dynamics of the fast DoFs, the perturbations, in Cauchy slices. As we already mentioned in the Introduction, the separation into slow/fast DoFs shows a natural association to bulk/boundary properties of the system once a hyperboloidal foliation of the Lorentzian manifold $M_2$ is considered~\cite{Gasperin:2021kfv}. In addition, the hyperboloidal slicing is particularly meaningful when studying properties of the system at null infinities. In our framework, it represents a powerful tool to connect bulk and boundary properties of the system. Indeed, we introduce in this section the necessary following step in the construction: the preservation of the integrable structure related to the slow DoFs in the hyperboloidal slicing. This sets the stage for future investigations analyzing the interplay with the boundary/asymptotic features of the system.

In this section we first review the hyperboloidal slicing for master equations and then show that it is invariant under scale transformations of the master function.
This feature, of interest by itself, turns out to be crucial in the study of how the bulk symmetries previously described in the Cauchy slices, leak in the hyperboloidal picture, while consistently preserving the bulk/boundary interpretation.

\subsection{Basics of the hyperboloidal slicing}

Let us first review the hyperboloidal slicing. The starting point is the following wave master equation:
\begin{equation}
\left( - \partial^2_t + \partial^2_{x} - V^{}_\ell \right) \phi = \mathcal{F} \,,
\label{general-master-equation}
\end{equation}
where we have used again the notation $x$ for the tortoise coordinate. Moreover, in this equation $V_\ell$ is the $\ell$-dependent potential and $\mathcal{F}$ is a generic source term.  Although we have not introduced a source term in the Cauchy-based formulation (it can only come from the existence of energy-momentum distributions entering at the perturbative order), we include it here for completitude and to show how it is affected by the different transformations that we are going to consider.

We can now implement the hyperboloidal compactification by making the following coordinate change:
\begin{eqnarray}
(t,x) \rightarrow (\tau,\xi) ~:~
\left\{
\begin{array}{lcl}
t  & = & \tau - h(\xi) \,,\\[2mm]
x  & = & g(\xi)  \,.
\end{array}
\right.   \,.
\label{hyperboloidal-compactification-coordinate-change}
\end{eqnarray}
Here, the height function $h(\xi)$ is such that surfaces with $\tau= \mathrm{const}$ intersect future null infinity $\scri^{+}$ and cross the horizon ($r=r_s$). The function $g(\xi)$, on the other hand, provides the compactification of the tortoise coordinate $x$ into the variable $\xi$ varying in a compact domain.
By applying this coordinate change to Eq.~\eqref{general-master-equation} and doing some algebra we can get to the following equation:
\begin{widetext}
\begin{eqnarray}
-\left[1-\left(\frac{h'}{g'}\right)^2\right] \partial^2_{\tau} \phi +
\frac{2h'}{(g')^2} \partial^2_{\tau \xi}\phi + 
\frac{1}{(g')^2} \partial^2_\xi\phi + 
\frac{1}{g'}\left(\frac{h'}{g'} \right)' \partial^{}_\tau\phi +
\frac{1}{g'}\left(\frac{1}{g'}\right)' \partial^{}_\xi \phi -
V^{}_\ell \phi = \mathcal{F} \,. 
\end{eqnarray}    
\end{widetext}
By introducing the following definition
\begin{equation}
\psi = \partial^{}_\tau \phi \,,   
\label{phi_tau}
\end{equation}
and isolating $\partial^{}_{\tau} \psi$ we can rewrite the previous equation as:
\begin{widetext}
\begin{eqnarray}
\partial^{}_{\tau} \psi =  \frac{1}{1-\left(\frac{h'}{g'}\right)^2} \left\{ 
\frac{2h'}{(g')^2} \partial^{}_{\xi} \psi + 
\frac{1}{(g')^2} \partial^2_\xi\phi + 
\frac{1}{g'}\left(\frac{h'}{g'} \right)' \psi +
\frac{1}{g'}\left(\frac{1}{g'}\right)' \partial^{}_\xi \phi -
V^{}_\ell \phi - \mathcal{F} \right\} \,.
\label{psi_tau}
\end{eqnarray}    
\end{widetext}
Equations~\eqref{phi_tau} and~\eqref{psi_tau} can be seen as evolution equations for $\phi$ and $\psi$ respectively.  Actually, they can be written in matrix form in the following way:
\begin{eqnarray}
\partial^{}_\tau \left(\begin{array}{c} \phi \\ \psi \end{array} \right) = i \mathbb{L} \left(\begin{array}{c} \phi \\ \psi \end{array} \right) + \left(\begin{array}{c} 0 \\ -\frac{\mathcal{F}}{1-\left(\frac{h'}{g'}\right)^2} \end{array} \right) \,,
\end{eqnarray}
where
\begin{eqnarray}
\mathbb{L} = \frac{1}{i} \left(  
\begin{array}{cc}
   0  &  1 \\[2mm]
\mathcal{L}^{}_1  &  \mathcal{L}^{}_2
\end{array}
\right)\,,
\label{L-structure}
\end{eqnarray}
and $\mathcal{L}_1$ and $\mathcal{L}_2$ are operators that act in the following way:
\begin{eqnarray}
\mathcal{L}^{}_1 \phi & = & \frac{1}{1-\left(\frac{h'}{g'}\right)^2} \left[ \frac{1}{(g')^2} \partial^2_\xi +  \frac{1}{g'}\left(\frac{1}{g'}\right)' \partial^{}_\xi  - V^{}_\ell \right]\phi \,, \qquad \\[2mm]
\mathcal{L}^{}_2 \psi & = & \frac{1}{1-\left(\frac{h'}{g'}\right)^2} \left[ \frac{2h'}{(g')^2} \partial^{}_{\xi} + \frac{1}{g'}\left(\frac{h'}{g'} \right)'\right] \psi \,.
\end{eqnarray}
These two operators can be written in a more convenient way. Let us start by  $\mathcal{L}_1$. We can write it in the following more familiar Sturm-Liouville form:
\begin{eqnarray}
\mathcal{L}^{}_1 \phi = \frac{1}{w(\xi)}\left[ \partial^{}_\xi\left(p(\xi)\partial^{}_\xi \right)  - q^{}_\ell(\xi) \right] \phi \,,
\end{eqnarray}
where
\begin{eqnarray}
\label{w}
w(\xi) & = & \frac{1}{g'}\left( g'^2 - h'^2 \right) \,, \\[2mm]
\label{p}
p(\xi) & = & \frac{1}{g'} \,, \\[2mm]
\label{q}
q^{}_\ell(\xi) & = & g' V^{}_\ell \,.
\end{eqnarray}
On the other hand, the action of $\mathcal{L}_2$ can be written as:
\begin{eqnarray}
\mathcal{L}^{}_2 \psi = \frac{1}{w(\xi)}\left[2\gamma(\xi)\partial^{}_\xi + \partial^{}_\xi\gamma(\xi)\right]\psi \,,
\end{eqnarray}
where the only new quantity is $\gamma(\xi)$, which is given by:
\begin{equation}
\gamma(\xi) = \frac{h'}{g'} \,.    
\end{equation}
There is an interesting feature of the structure of this operator which provides a different perspective on the separation into slow and fast DoFs. It was shown in Refs.~\cite{Gasperin:2021kfv,Jaramillo:2020tuu} that the non-self-adjointedness of the operator $\mathbb{L}$ is associated to the operator $\mathcal{L}_2$, which account for the dissipative character of the system at $\scri^+$ and the horizon ($r=r_s$), i.e. the asymptotic boundaries. Taking this into account, the operator $\mathcal{L}_1$ is associated to “bulk” properties of the system while $\mathcal{L}_2$ to “boundary” properties. This supports the conclusion that, within perturbation theory, the bulk properties of the system are expressed by the slow DoFs, in the sense explained above (see Refs.~\cite{Gasperin:2021kfv,Jaramillo:2020tuu} for more detailed arguments).

\subsection{Scale covariance in hyperboloidal slicing}

After reviewing the hyperboloidal compactification for the general master equation in Eq.~\eqref{general-master-equation}, we are going to show a novel general covariance of the hyperboloidal slicing approach under the following transformation of the wave function [in the spirit of the transformation in Eq.~\eqref{conformal-transformation-psi-x}]: 
\begin{equation}
\phi(\tau,\xi) = \Omega(\xi) \tilde{\phi}( \tau,\xi) \,.
\label{transformation-wave-function}
\end{equation}

Then, starting from Eq.~\eqref{general-master-equation}, introducing the coordinate change that implements the hyperboloidal slicing [transformations in Eq.~\eqref{hyperboloidal-compactification-coordinate-change}], introducing the transformation of the wave function [Eq.~\eqref{transformation-wave-function}], and after some algebra, we can reach the following expression:
\begin{widetext}
\begin{eqnarray}
&& -\Omega\left[1-\left(\frac{h'}{g'}\right)^2\right] \partial^2_{\tau} \tilde{\phi} +
\frac{2h'\Omega}{(g')^2} \partial^2_{\tau \xi} \tilde{\phi} + 
\frac{\Omega}{(g')^2} \partial^2_\xi \tilde{\phi} + 
\frac{1}{g'}\left[ \frac{h'\Omega'}{g'} + \left(\frac{h'\Omega}{g'} \right)' \right] \partial^{}_\tau \tilde{\phi} +
\frac{1}{g'}\left[\frac{\Omega'}{g'} + \left(\frac{\Omega}{g'}\right)'\right] \partial^{}_\xi \tilde{\phi} \nonumber \\[2mm]
& & + \left[ \frac{1}{g'}\left(\frac{\Omega'}{g'}\right)' - \Omega V^{}_\ell \right] \tilde{\phi} = \mathcal{F} \,. 
\label{hyperboloidal-general-transformation}
\end{eqnarray}    
Introducing the functions defined in Eqs.~\eqref{w},~\eqref{p} and~\eqref{q}, we can rewrite this as
\begin{eqnarray}
 \left\{-\Omega\left[1-\gamma^2\right] \partial^2_{\tau}  +
2p\gamma\Omega\, \partial^2_{\tau \xi}  + 
p^2\Omega\, \partial^2_\xi  + 
p\left[ \gamma\Omega'+ \left(\gamma\Omega \right)' \right] \partial^{}_\tau +
p\left[p\Omega' + \left(p\Omega \right)'\right] \partial^{}_\xi  
 + \left[ p\left(p\Omega' \right)' - \Omega V^{}_\ell \right]  \right\}\tilde{\phi} = \mathcal{F} \,.  \qquad
\end{eqnarray}    
\end{widetext}
Following the same steps as before, and introducing a new variable for the time derivative of $\tilde{\phi}$
\begin{equation}
\tilde{\psi} = \partial^{}_\tau \tilde{\phi} \,,   
\label{dPhidtau}
\end{equation}
the wave equation can be cast into the following form:
\begin{eqnarray}
\nonumber
    \Big[-\partial^{}_\tau  &+& \frac{1}{\tilde{w}}\left( 2\tilde{\gamma}\partial^{}_\xi +\partial^{}_\xi \tilde{\gamma} \right) \Big]\tilde{\psi} \\
    &+& \frac{1}{\tilde{w}}\left[\partial^{}_\xi\left(\tilde{p}\partial^{}_\xi\right) -\frac{\tilde{V}}{\tilde{p}} \right]\tilde{\phi} = \mathcal{F}
    \,,
\end{eqnarray}
where we have defined the following scaled quantities
\begin{eqnarray}
\tilde{w}(\xi) & = & \Omega^2(\xi) w(\xi)\,, \\[2mm]
\tilde{p}(\xi) & = & \Omega^2(\xi) p(\xi) \,, \\[2mm]
\tilde{\gamma}(\xi) & = & \Omega^2(\xi) \gamma(\xi) \,,
\end{eqnarray}
and the potential $\tilde{V}$ is defined as
\begin{eqnarray}
    \tilde{V} =   \Omega^3p\left[\frac{\Omega V}{p} -\partial^{}_\xi\left(p\partial^{}_\xi \Omega\right)\right] \,.
\end{eqnarray}
Rewriting the master equation in matrix form gives
\begin{eqnarray}
\partial^{}_\tau \left(\begin{array}{c} \tilde{\phi} \\[2mm] \tilde{\psi} \end{array} \right) = i \tilde{\mathbb{L}} \left(\begin{array}{c} \tilde{\phi} \\[2mm] \tilde{\psi} \end{array} \right) + \left(\begin{array}{c} 0 \\[2mm] -\frac{\mathcal{F}}{\Omega\left(1-\gamma^2\right)} \end{array} \right) \,,
\end{eqnarray}
where
\begin{eqnarray}
\mathbb{\tilde{\mathbb{L}}
} = \frac{1}{i} \left(  
\begin{array}{cc}
   0  &  1 \\[2mm]
\tilde{\mathcal{L}}^{}_1  &  \tilde{\mathcal{L}}^{}_2
\end{array}
\right)\,,
\label{L-structure-covariant}
\end{eqnarray}
and $\tilde{\mathcal{L}}^{}_1$ and $\tilde{\mathcal{L}}^{}_2$ are operators that act in the following way:
\begin{eqnarray}
\tilde{\mathcal{L}}^{}_1 \tilde{\phi} & = & \frac{1}{\tilde{w}}\left[\partial^{}_\xi\left(\tilde{p}\partial^{}_\xi\right) -\frac{\tilde{V}}{\tilde{p}} \right]\tilde{\phi}  \,, \\[2mm]
\tilde{\mathcal{L}}^{}_2 \tilde{\psi} & = & \frac{1}{\tilde{w}}\left( 2\tilde{\gamma}\partial^{}_\xi +\partial^{}_\xi \tilde{\gamma} \right)\tilde{\psi} \,.
\end{eqnarray}
This shows the complete covariance of the hyperboloidal approach under transformations of the master function by a scale factor $\Omega$, as indeed the $\tilde{\mathbb{L}}$ operator preserves the same structure as in the ordinary case~\eqref{L-structure}, with the operator $\tilde{\mathcal{L}}^{}_1$ having a Sturm-Liouville form. This non-trivial result introduces an apparently arbitrary freedom which may turn out to be useful in several application, including numerical calculations.

\subsection{Conformal hyperboloidal slicing}

With the purpose of making contact with the bulk symmetries in Cauchy slices and understand whether such symmetric structure is present also in the hyperboloidal slices, we are going to fix $\Omega$ by cancelling the term containing $\partial^{}_\xi \tilde{\phi}$ as we have already done it in Sec.~\ref{Sec:conformal-cauchy} for the case of Cauchy slices. Looking back at Eq.~\eqref{hyperboloidal-general-transformation}, this vanishing condition implies
\begin{eqnarray}
\frac{\Omega'}{\Omega} = \frac{g''}{2 g'}  \,,  
\end{eqnarray}
which can be solved for $\Omega$ to give
\begin{eqnarray}
\Omega(\xi) =  \Omega^{}_o \sqrt{g'(\xi)}  =  \Omega^{}_o \, p_{}^{-1/2} (\xi) \,,  \label{omega-expression}
\end{eqnarray}
where $\Omega_o$ is a constant.  With this, the general wave master equation becomes:
\begin{widetext}
\begin{eqnarray}
&& -\left[1-\left(\frac{h'}{g'}\right)^2\right] \partial^2_{\tau} \tilde{\phi} +
\frac{2h'}{(g')^2} \partial^2_{\tau \xi} \tilde{\phi} + 
\frac{1}{(g')^2} \partial^2_\xi \tilde{\phi} + 
\frac{h''}{g'^2} \partial^{}_\tau \tilde{\phi} +
\left[\frac{1}{2g'^2} \mathcal{S}(g) - V^{}_\ell \right] \tilde{\phi} = \frac{\mathcal{F}}{\Omega} \,. 
\end{eqnarray}    
\end{widetext}
Starting again from Eq.~\eqref{hyperboloidal-general-transformation} and introducing the first order (in time) field $\tilde{\psi}$ [see Eq.~\eqref{dPhidtau}], we get to the following master wave equation
\begin{widetext}
\begin{eqnarray}
\partial^{}_\tau\tilde{\psi} = \frac{1}{1-\left(\frac{h'}{g'}\right)^2} \left\{ 
\frac{2h'}{(g')^2} \partial^{}_{\xi} \tilde{\psi} + 
\frac{1}{(g')^2} \partial^2_\xi \tilde{\phi} + 
\frac{h''}{g'^2} \tilde{\psi} +
\left[\frac{1}{2g'^2} \mathcal{S}(g) - V^{}_\ell \right] \tilde{\phi} - \frac{\mathcal{F}}{\Omega}\right\} \,. 
\end{eqnarray}    
\end{widetext}
We can again perform a reduction to a first order system of equations:
\begin{eqnarray}
\partial^{}_\tau \left(\begin{array}{c} \tilde{\phi} \\[2mm] \tilde{\psi} \end{array} \right) = i \mathbb{L} \left(\begin{array}{c} \tilde{\phi} \\[2mm] \tilde{\psi} \end{array} \right) + \left(\begin{array}{c} 0 \\[2mm] -\frac{\mathcal{F}/\Omega}{1-\left(\frac{h'}{g'}\right)^2} \end{array} \right) \,,
\end{eqnarray}
where $\mathbb{L}$ has the same structure as in Eq.~\eqref{L-structure}, where $\mathcal{L}_1$ and $\mathcal{L}_2$ now act as follows:
\begin{eqnarray}
\mathcal{L}^{}_1 \tilde{\phi}  =  \frac{1}{1-\left(\frac{h'}{g'}\right)^2} \left\{ \frac{1}{(g')^2} \partial^2_\xi +  
\left[\frac{1}{2g'^2} \mathcal{S}(g) - V^{}_\ell \right] \right\} \tilde{\phi} \,, 
\end{eqnarray}
\begin{equation}
\mathcal{L}^{}_2 \tilde{\psi} = \frac{1}{1-\left(\frac{h'}{g'}\right)^2} \left[ \frac{2h'}{(g')^2} \partial^{}_{\xi} + \frac{h''}{g'^2} \right] \tilde{\psi} \,.
\label{L2-conformal}
\end{equation}
The action of the  operator $\mathcal{L}_1$ can be rewritten as follows:
\begin{equation}
\mathcal{L}^{}_1 \tilde{\phi} = \frac{1}{g'^2 - h'^2}\left(\partial^2_\xi - \tilde{V}^{}_\ell \right)\tilde{\phi}\,,
\end{equation}
where the new potential $\tilde{V}_\ell$ involves again the Schwarzian derivative (see below):
\begin{equation}
\tilde{V}^{}_\ell = g'^2\, {V}^{}_\ell - \frac{1}{2} \mathcal{S}(g) \,.
\label{potential-schwarzian-g}
\end{equation}
And the action of the operator $\mathcal{L}_2$ can be written as
\begin{equation}
\mathcal{L}^{}_2 \tilde{\psi} = \frac{1}{g'^2 - h'^2}\left\{h'\,,\,\partial^{}_\xi \right\}\tilde{\psi}\,,
\end{equation}
where $\{\cdot , \cdot \}$ denotes the anti-commutator of operators:
\begin{equation}
\left\{A\,,\,B\right\}\Psi \equiv A(B(\Psi)) + B(A(\Psi)) \,.
\end{equation}
Using the following interesting property of the Schwarzian:
\begin{equation}
\mathcal{S}\left(g^{-1}(\xi)\right) = \mathcal{S}\left(g(\xi)\right)  \,,
\end{equation}
we observe that the same conformal/Virasoro structure that was present in the Cauchy slices, is reproduced in the hyperboloidal setting, namely, the potential transforms as the holomorphic components of the stress-energy momentum tensor.

\section{Conclusions and Discussion}\label{Conclusions-and-Discussion}

In this work, we have investigated integrable hidden structures present in the equations governing the dynamics of perturbed BHs. The study reveals an important and novel connection of the dynamics with the properties of integrable systems. From a more global point of view, this work constitutes a step forward towards a longer term objective: the study of structurally stable features in BBH dynamics by adopting an integrability approach based on the effective separation into slow/bulk and fast/boundary DoFs in the description of BBH systems, as proposed in Ref.~\cite{Jaramillo:2023day} (see also \cite{Jaramillo:2022oqn,Jaramillo:2022kuv,Jaramillo:2024nvr}). In this sense, BHPT offers a perfect arena to start studying some of these questions in a simplified setting in which this separation is clearer. On the other hand, however, it also suggests the possibility of broader applications, not restricted to the dynamics of BBHs. One of the main goals of such program is the study of the interplay between the “bulk” symmetries described here and the asymptotic symmetries at null infinity (the BMS group of symmetries or, possibly, an appropriate extension), an interplay that admits a natural set up in the hyperboloidal approach to BH dynamics ~\cite{Gasperin:2021kfv}. The present work can then be seen as representing an important and necessary milestone in that direction, by establishing the bulk integrable structures hidden in the dynamics of the slow DoFs.

In previous works~\cite{Lenzi:2021njy,Lenzi:2022wjv}, it has been shown that the infinite hierarchy of KdV conserved quantities can be associated to the description of gravitational BH perturbations, in Cauchy slices, as a consequence of the isospectral character of KdV deformations of the spectral problem. In this work, this hidden symmetry has been expanded by exploiting the bi-Hamiltonian structure of the KdV equation. On the one hand, the Magri brackets, realizing the second Hamiltonian KdV flow, are associated to the isospectral (KdV) deformations of the master equations as they provide the KdV Hamiltonian flow.
On the other hand, the Magri brackets are known to be the classical representation of the Virasoro algebra~\cite{Gervais:1985fc,GERVAIS1985279}. We have introduced this structure in the description of BHs by identifying the BH potential response to gravitational perturbations with (one component of) the energy momentum tensor of a $2d$ CFT with central charge $c=-6$. 
This suggests an interesting twofold interpretation of the BH potential. First, when we KdV-deform the potential, it can be seen as a class of solitonic initial data for the KdV dynamics given by the Magri brackets.
If instead, we analytically continue the BH potential into the complex plane and take its Laurent expansion, then it defines the generators of a Virasoro algebra, again through the use of the Magri brackets. Moreover, the infinitesimal conformal transformation of the BH potential is also given by the Magri brackets between the BH potential and the conformal charge. The exponentiation of the infinitesimal transformation shows that the BH potential transforms as a quasi-primary field, whose deviation from being a primary field is provided by the Schwarzian derivative of the coordinate transformation. We then have explored how this affects the master equation describing the dynamics of BH perturbations. We can therefore identify a “triangle” of mathematical structures, which underlies the symmetries we have just described, with the interplay KdV-Virasoro-Schwarzian derivative~\cite{Semenov_r-matrix_Approach}, which provides a guiding map between KdV isospectral deformations and conformal symmetries in the equations describing BH perturbations. 

In addition, we have studied the uplift of these structures into a formulation of the BH perturbations that uses a hyperboloidal foliation of spacetime instead of the Cauchy slicings, traditionally used in integrability settings. This provides a relevant leap since, within the hyperboloidal framework, the bulk character of the KdV integrable structures is more evident and, at the same time, it allows us to prepare the ground for studying the connection with asymptotic structures at null infinity. As an intermediate step, we have shown that the master equations describing BH perturbations, formulated on a hyperboloidal slicing, are covariant with respect to an additional freedom added to the system, namely scaling transformations of the master function. This represents a novel feature of the hyperboloidal approach that has not been explored before and whose consequences should not be overlooked, as we further discuss them in the next paragraph. Starting from this set up, if we restrict the class of scaling transformations so that the master function transforms as a primary field with respect to coordinate transformations of the spatial coordinate, then we recover the conformal transformation of the BH potential as a quasi-primary field in terms of the Schwarzian derivative, as in the case of a Cauchy slicing. This shows that the Virasoro algebra governing the “bulk” properties of the system is preserved in a hyperbolidal slicing. There is however a caveat in doing this, which is related to the fact that isospectrality and covariance properties of differential operators are not always simultaneously guaranteed~\cite{Zuber:1991vs,DiFrancesco:1990qr}. Indeed, the covariance under coordinate changes of differential operators is related to the so-called $W$-algebras, an extension of the Virasoro algebra~\cite{Belavin:1984vu,Zamolodchikov:1985wn, DiFrancesco:1990qr,Dickey:1997wia}. However, this is beyond the scope of this paper and should be dealt with in future research.

The results of this work suggest a number of developments in different directions. First, the novel transformation properties introduced in the hyperboloidal framework, both the covariance and the conformal ones, can in principle provide useful tools in the numerical study of the ringdown emission in BBHs. Indeed, the covariance of master equations under general scalings of the master functions introduces an additional freedom that allows in principle to perform simulations using a rescaled master function in a way that can be explored to address and by-pass potential numerical problems associated with the functional form of the straightforwardly obtained equations. Moreover, the transformation of the BH potential according to the Schwarzian derivative introduces at least two other interesting possibilities. On the one hand, we can freely perform coordinate changes given by any M\"obius transformation and this will leave the BH potential invariant. On the other hand, it opens the door to the intriguing possibility of performing a “conformal bridge” transformation~\cite{Inzunza:2019sct,Achour:2021dtj} which, in some cases, has been shown to relate confined systems to free ones. 
Another future direction that this study suggests is a completion of this description of the isospectral character of “bulk” symmetries in a hyperboloidal slicing. This implies to properly account for the previously mentioned interplay between covariance and isospectral properties of differential operators~\cite{Zuber:1991vs}, in an attempt to find the Lax pair associated with the infinitesimal generator of the dynamics in this setting, namely the hyperboloidal operator $\mathbb{L}$. We expect this issue to be intimately related to the asymptotic/boundary structures.

Finally, the relevance of these hidden symmetries in addressing some important physical aspects of BH spacetimes and their dynamics should not be discarded. First, BHs are known to have peculiar tidal properties as their Love numbers, measuring the tidal deformability, identically vanish~\cite{Binnington:2009bb,Damour:2009vw}, thus implying the impossibility to tidally deform a BH. This result has recently triggered a number of works trying to relate such property to underlying hidden symmetries~\cite{Hui:2020xxx,Charalambous:2021mea,Charalambous:2021kcz,Hui:2021vcv, BenAchour:2022uqo}. On the other hand, hidden (conformal) symmetries have been shown to provide an algebraic tool to estimate the BH QNMs in non-asymptotically flat spacetimes (see e.g.~\cite{Birmingham:2001pj,Bertini:2011ga}). The novel symmetries in the dynamics of perturbed asymptotically flat spacetimes found in this work may provide interesting insights into the behaviour of BH QNMs, in particular their spectral (in)stability.

%
\begin{acknowledgments}
We warmly thank  Badri Krishnan for the shared efforts to define and push the `universality from integrability' program to BBH dynamics. We also thank Carlos Barcel\'o, J\'er\'emy Besson, Piotr Bizo\'n,  Edgar Gasper\'\i n, Vincent Lam, Juan A. Valiente-Kroon, Corentin Vitel and Mikhail Semenov Tian-Shansky for enlightening discussions.
ML and CFS are supported by contracts PID2019-106515GB-I00 and PID2022-137674NB-I00 from MCIN/AEI/10.13039/501100011033 (Spanish Ministry of Science and Innovation) and 2017-SGR-1469 and 2021-SGR-01529 (AGAUR, Generalitat de Catalunya).  
ML is also supported by Juan de la Cierva contract FJC2021-047289-I funded by program MCIN/AEI/10.13039/501100011033 (Spanish Ministry of Science and Innovation) and by NextGenerationEU/PRTR (European Union).
This work has also been partially supported by the program \textit{Unidad de Excelencia Mar\'{\i}a de Maeztu} CEX2020-001058-M (Spanish Ministry of Science and Innovation). JLJ is supported by the program "Investissements d’Avenir" through project ISITE-BFC (ANR-15-IDEX-03), the ANR "Quantum Fields interacting with Geometry" (QFG) project (ANR-20-CE40-0018- 02), the EIPHI Graduate School (ANR-17- EURE-0002) and the Spanish FIS2017-86497-C2-1 project (with FEDER contribution).
\end{acknowledgments}

\appendix

\section{Hamiltonian Formalism in the Infinite–Dimensional Case}
\label{Hamiltonian-Formalism}

In this case we have a continuous of dynamical variables, here labeled by the coordinate $x$.
The changes with respect to the finite-dimensional case are that trajectories are replaced by smooth functions $V(\sigma,x)$ and phase space functions are replaced by functionals. A functional of $V$ is defined via integration as follows:
\begin{equation}
F[V] = \int^{\infty}_{-\infty} dx\,f(V,V^{}_{,x}, V^{}_{,xx}, \ldots ) \,. 
\end{equation}
Functional (Fr\`echet) derivatives are defined as~\cite{Olver1998:pj}:
\begin{eqnarray}
\frac{\delta F}{\delta V(x)} & = & \sum_{n=0}^\infty (-1)^n \frac{\partial^n}{\partial x^n}
\frac{\partial f}{\partial V^{(n)}} \nonumber \\
& = & \frac{\partial f}{\partial V} - \frac{\partial}{\partial x}\frac{\partial f}{\partial V^{}_{,x}} + \frac{\partial^2}{\partial x^2}\frac{\partial f}{\partial V^{}_{,xx}} + \ldots  \,,
\end{eqnarray}
where $V^{(n)} = d^nV(x)/dx^n$ and one has to take into account that
\begin{equation}
\frac{\delta V(x)}{\delta V(x')} = \delta(x-x') \,,
\end{equation}
where $\delta(x)$ denotes the Dirac delta distribution.  We can build  Poisson brackets as
\begin{equation}
\left\{ F , G \right\} = \int^{\infty}_{-\infty}\!\!\!\! dx\! \int^{\infty}_{-\infty} \!\!\!\! dx'  \omega(x,x',V) \frac{\delta F}{\delta V(x)} \frac{\delta G}{\delta V(x')} \,,
\end{equation}
where $\omega(x,x',V)$, the symplectic form, has to be such that the Poisson bracket is antisymmetric.  A common choice is~\cite{gardner1971korteweg} 
\begin{equation}
\omega(x,x',V) = \frac{1}{2}\left( \partial^{}_x - \partial^{}_{x'} \right) \delta(x-x') \,.
\label{GZF-form}
\end{equation}
Since the partial derivative operator $\partial_x$ is anti-self-adjoin with respect to the inner scalar product
\begin{equation}
\left( V, W \right) = \int^{\infty}_{-\infty} dx\,V(x) W(x)\,,
\end{equation}
the Poisson bracket can be written as:
\begin{equation}
\left\{ F , G \right\}^{}_{\rm GZF} =   \int^{\infty}_{-\infty} dx\, \frac{\delta F}{\delta V(x)} \mathcal{D}\frac{\delta G}{\delta V(x)} \,,
\label{GZF-bracket}
\end{equation}
where
\begin{equation}
\mathcal{D} = \frac{\partial}{\partial x} \,.    
\end{equation}
The Poisson bracket in Eq.~\eqref{GZF-bracket} is sometimes referred to as the Gardner-Zakharov-Faddeev bracket~\cite{gardner1971korteweg,Zakharov:1971faa}.
In this way, Hamilton's equations can be written as follows:
\begin{equation}
\frac{\partial V}{\partial\sigma}= \left\{ V , \mathcal{H}[V] \right\}^{}_{\rm GZF}  = \mathcal{D} \frac{\delta \mathcal{H}[V]}{\delta V(x)} \,.
\end{equation}
A different choice of symplectic form was introduced by Magri~\cite{Magri:1977gn}
\begin{eqnarray}
\omega(x,x',V) & =& \left[ - \frac{1}{2}\left(\partial^{3}_x - \partial^{3}_{x'}\right) \right. \nonumber \\
& + & \left. 2\left(V(x)\partial^{}_x - V(x')\partial^{}_{x'}\right) \right] \delta(x-x') \,.
\label{Magri-form}
\end{eqnarray}
This leads to the following Poisson bracket:
\begin{equation}
\left\{ F , G \right\}^{}_{\rm M} = \int^{\infty}_{-\infty} dx\, \frac{\delta F}{\delta V(x)}\mathcal{E}^{}_V \frac{\delta G}{\delta V(x)} \,,
\label{Magri-bracket}
\end{equation}
where $\mathcal{E}^{}_V$ is defined in Eq.~\eqref{magri-operator}.

%

\end{document}